\documentclass[%
 reprint,
superscriptaddress,
 amsmath,amssymb,
 aps,
pra,
]{revtex4-1}

\usepackage{graphicx}
\usepackage{dcolumn}
\usepackage{bm}
\usepackage{comment}

\newcommand{\eref}[1]{Eq.~(\ref{#1})}
    
\begin{document}

\preprint{APS/123-QED}

\title{
Population kinetics of many-electron atoms in ionizing plasmas studied using a continuous collisional radiative model
} 

\author{Akira Nishio}
 \affiliation{%
Department of Mechanical Engineering and Science,
Graduate School of Engineering, Kyoto University,
Kyoto 615-8540, Japan}

\author{Julian C. Berengut}%
\affiliation{%
School of Physics, University of New South Wales, New South Wales 2052, Australia}

\author{Masahiro Hasuo}%
\affiliation{%
Department of Mechanical Engineering and Science,
Graduate School of Engineering, Kyoto University,
Kyoto 615-8540, Japan} 
\email{fujii@me.kyoto-u.ac.jp}

\author{Keisuke Fujii}
 \affiliation{%
Department of Mechanical Engineering and Science,
Graduate School of Engineering, Kyoto University,
Kyoto 615-8540, Japan} 
\email{fujii@me.kyoto-u.ac.jp}

\date{\today}

\begin{abstract}
Collisional--radiative (CR) models based on \textit{ab initio} atomic structure calculation have been utilized over 20 years to analyze many-electron atomic and ionic spectra. 
Although the population distribution of the excited states in plasmas and their emission spectra are computed using the CR models, systematic and analytical understanding of the population kinetics are still lacking.
In this work, we present a reduced model of the population dynamics in many-electron atomic ions, in which we approximate the dense energy structure of complex many-electron atoms by a continuum, a continuous CR model (CCRM).
Using this simplification, we show an analytical population distribution of many-electron atoms in plasmas and its electron-density and temperature dependence.
In particular, the CCRM shows that the population distribution of highly excited states of many-electron atoms in plasmas resembles a Boltzmann distribution but with an effective excitation temperature.
We also show the existence of three typical electron-density regions and two electron-temperature regions where the parameter dependence of the excitation temperature is different.
Analytical representations of the effective excitation temperature and the boundaries of these phases are also presented.
\end{abstract}

\maketitle


\section{Introduction}

The spectra of many-electron atomic ions can be seen in various optically thin plasmas.
In the stellar atmosphere, neutral and singly charged iron (Fe) are the dominant components in the absorption spectra in terms of the number of lines~\cite{Tousey1988}.
Many of Fe absorption lines have been identified to study the stellar atmosphere~\cite{Nave1994, Castelli2010, Peterson2017}.
Highly charged tin and actinoide ions play an important role in realizing ultraviolet light sources~\cite{Osullivan2015,Suzuki2012,Torretti2020}, in which quasi-continuum emission in laser-produced plasmas is used.
Since the radiative power should be concentrated into a particular energy region for the commercial light source realization, many works have been carried out to understand the population dynamics in plasmas~\cite{Osullivan2015,Torretti2020}.
In fusion tokamak plasmas, highly charged tungsten ions convert electron kinetic energy to strong radiation and therefore need to be controlled~\cite{Putterich2008,Murakami2015}.
The thermalization process of the nuclei in kilonovae, which has recently been probed from the emission of neutral transition metals, is yet to be understood~\cite{Pian2017,Tanaka2018}.

The collisional--radiative (CR) model is the key tool to study the population kinetics of many-electron atoms in plasmas and their emission and absorption spectra.
This model solves the steady state equation of the excited state population of ions by taking into account the rates of elementary processes in plasmas.
In order to perform accurate predictions, accurate atomic data are required, i.e., energy levels and transition rates of many elementary processes, including electron-impact excitations and spontaneous transitions.
Therefore, many works have been dedicated to develop and improve \textit{ab initio} calculation of these atomic data~\cite{Cowan1981,Gu2008,amusia1997computation,Bar-Shalom2001,Hill2018}.

Although this first-principles approach has been successful in many cases~\cite{TheVenin1999,Dodin2014,Torretti2020,Murakami2015,RalchenkoEditor2016}, in order to understand properties of the simulated population dynamics and spectra, different approaches are required.
One of such properties is the analytical criterion of the local thermal equilibrium (LTE), where the population distribution is perfectly represented by the Boltzmann's distribution.
Some criteria have been reported for hydrogen-like ions by several authors \cite{Bates1962, Griem, Huddlestone1965, Fujimoto1973,Fujimoto1990} and they have been also applied for a-few-electron atoms, however, the applicability to many-electron atoms is not yet clear.

Another property of the population dynamics is the Boltzmann's-like population distribution. 
Busquet et al proposed the effective ionization temperature to describe the ion density abundance in laser produced plasmas~\cite{Busquet1993}.
This concept has been extended to represent the population distribution in excited states, and has been confirmed both experimentally and numerically even in far condition from the LTE \cite{Blancard2003,Li2010,Hansen2006,Bauchea}.
It has been also theoretically explained that the population in each configuration \cite{Fournier2000, Fournier2000, Bauche-Arnoult2001} and in each super-configuration \cite{Bauchea} become similar to a Boltzmann's distribution.
Furthermore, the parameter dependence of the effective excitation temperature, which determines the slope of the Boltzmann's-like distribution, has been estimated based on the two-level model \cite{Hansen2006}.
This property of population dynamics has been utilized to skip the heavy collisional-radiative model computation in radiation-fluid simulation \cite{Busquet1993,Bauchea}.

However, it is not clear whether this property still holds in complex many-electron atoms, where the \textit{configuration} is not well defined because of a strong wavefunction mixing.
The applicability of the two-level model for the excitation temperature estimation is not clear either, as many-electron atoms have huge number of levels and the population transfer among them may be significant.
In this work, we show that the Boltzmann's-like distribution also holds for many-electron atoms.
We also present an analytical LTE criterion for many-electron atoms based on the statistical approximation of their atomic structure.

For this purpose, We approximate the dense energy levels of many-electron atoms by a continuum and the transition between them by kernel-integrations.
We call this model as a continuous CR model (CCRM).
In this CCRM, only two atomic parameters are used to represent the population kinetics: the energy scale of the level density growth; and another energy scale that describes the decay of transition strengths.
With this model, the dependence of the excitation temperature on electron density ($n_\mathrm{e}$) and temperature ($T_\mathrm{e}$) are studied using the CCRM, revealing the existence of three typical $n_\mathrm{e}$ regions and two $T_\mathrm{e}$ regions.
The LTE criterion for the many-electron atoms are also derived.
Furthermore, it is shown that in low $T_\mathrm{e}$ regions, the excitation temperature becomes almost $T_\mathrm{e}$ even in low $n_\mathrm{e}$ plasmas.
This indicates much wider applicability of the Boltzmann's method, which is a well-known method to estimate $T_\mathrm{e}$ values from the emission spectra in high $n_\mathrm{e}$ plasmas, as well as the new temperature diagnostics based on the line intensity statistics~\cite{Fujii2020}.

It is also found that the above properties of many-electron atoms are similar to those derived from two-level systems \cite{Hansen2006}, but its energy interval of the two-level system should be replaced by the energy scale related to the level density of the many-electron atom.
We also discuss this difference, which reflects the effect of dense energy levels.

In section~\ref{sec:crmodel}, we briefly describe the principle of the CR model and show simulation results obtained using an \textit{ab initio} calculation code for several many-electron atomic ions.
In section~\ref{sec:model}, we present our CCRM to study the population kinetics of many-electron atoms and compare it with the \textit{ab initio} simulation result.
In section~\ref{sec:discussions}, we discuss its parameter dependence.

\section{CRM for Many-Electron Atoms\label{sec:crmodel}}
\subsection{Principle of the CR Model}
The CR model is a balance equation of the population of excited states in plasmas.
In optically thin plasmas, the dominant excitation/de-excitation processes are radiative decay and electron-impact excitation and de-excitation.
The temporal evolution of the population in an excited state $p$ can be written as
\begin{equation}
\label{eq:crm}
\frac{\mathrm{d} n_p}{\mathrm{d}t} = 
\mathcal{A}_p^\mathrm{in} 
- \mathcal{A}_p^\mathrm{out} 
+ \mathcal{C}_p^\mathrm{in} 
- \mathcal{C}_p^\mathrm{out}
+ \mathcal{F}_p^\mathrm{in} 
- \mathcal{F}_p^\mathrm{out},
\end{equation}
where $\mathcal{A}_p^\mathrm{in}$ and $\mathcal{A}_p^\mathrm{out}$ are the population influx from upper levels to state $p$ and the outflux from state $p$ to lower levels by spontaneous decay, respectively.
Similarly, $\mathcal{C}_p^\mathrm{in}$ and $\mathcal{C}_p^\mathrm{out}$ are the influx and the outflux by electron-impact excitation, respectively, and $\mathcal{F}_p^\mathrm{in}$ and $\mathcal{F}_p^\mathrm{out}$ are the influx and the outflux by electron-impact de-excitation, respectively.
Except for extreme cases, the time scale of the excited state population is very short compared with that of the bulk plasma parameters (e.g., $n_\mathrm{e}$ and $T_\mathrm{e}$).
Therefore, the steady state of the population of excited states can be assumed, $\mathrm{d} n_p / \mathrm{d}t = 0$.

Many other elementary processes can be included in \eref{eq:crm}, such as ionization, recombination, photoionization, and photoexcitation. For highly charged ions, dielectronic recombination and auto-ionization may be important.
However, for simplicity, we mainly focus on electron-impact excitation, de-excitation, and spontaneous decay in this work (excluding an \textit{ab initio} calculation FAC used as a reference).

The excited state population is divided into two components, i.e., ionizing and recombining plasma components, depending on whether the population contributions of the ground state or the next ionized stage are dominant~\cite{Fujimoto}.
However, in this work, we neglect the ionization and recombination processes, and therefore only consider the ionizing plasma component.

Each term in \eref{eq:crm} is explicitly written as follows:
\begin{align}
    \notag
    \mathcal{A}_p^\mathrm{in} &= \sum_{q > p} A_{p \gets q} n_q, & 
    \mathcal{A}_p^\mathrm{out} &= \sum_{q < p} A_{q \gets p} n_p\\
    \notag
    \mathcal{C}_p^\mathrm{in} &= \sum_{q < p} C_{p \gets q} n_\mathrm{e} n_q, & 
    \mathcal{C}_p^\mathrm{out} &= \sum_{q > p} C_{q \gets p} n_\mathrm{e} n_p\\
    \label{eq:crm_each_term}
    \mathcal{F}_p^\mathrm{in} &= \sum_{q > p} F_{p \gets q} n_\mathrm{e} n_q, & 
    \mathcal{F}_p^\mathrm{out} &= \sum_{q < p} F_{q \gets p} n_\mathrm{e} n_p
\end{align}
where $A_{p \gets q}$ is the spontaneous transition rate from $q$ to $p$ states, $C_{p \gets q}$ is the electron-impact excitation rate coefficient from $q$ state to $p$ state, and $F_{p \gets q}$ is the de-excitation rate coefficient.
Here, $\sum_{q<p}$ and $\sum_{q>p}$ indicate the summation over states $q$ with higher and lower excited energies than the energy of $p$, respectively.
Then, Equation (\ref{eq:crm}) becomes a linear equation of $n_p$, which can be solved if we know all the rates.

The radiative transition rate $A_{p \gets q}$ is related to the line strength $S_{pq}$ between $p$ and $q$ states.
The transition rate by electric dipole transitions, which are almost always dominant, can be written as 
\begin{equation}
    \label{eq:radiative_rate}
    A_{p \gets q} = \gamma \frac{1}{g_q}\omega_{pq}^3 S_{pq}
\end{equation}
with
\begin{equation}
    \label{eq:gamma}
    \gamma = \frac{4}{3} \alpha^4 \frac{c}{a_0} \frac{1}{E_\mathrm{H}^3}\frac{1}{e^2 a_0^2}
\end{equation}
where $\omega_{pq} = E_q - E_p$ is the energy difference between states $p$
and $q$, $g_q$ is the statistical weight of state $q$, $E_\mathrm{H} \approx$ 27.2 eV is the Hartree energy, $\alpha$ is the fine structure constant, $e$ is the elementary charge, $a_0$ is the Bohr radius, and $c$ is the light speed.

The electron-impact excitation rate coefficient is more complicated.
Various methods of its approximation are available, such as Born method, close-coupling method~\cite{Grant1974}, distorted-wave approximation~\cite{Guo-xin1996,Eissner1998}, convergent close-coupling method~\cite{Bray1992,Bray1992a}, and $R$-matrix method~\cite{Quigley1998,Burke1993}.
Besides such sophisticated methods, it has been known that the rate coefficients have a nearly proportional dependence to the line strength between the corresponding level pair \cite{Mewe1972, Fujimoto, griem_1997},
\begin{equation}
    \label{eq:excitation_rate}
    C_{p \gets q} \approx
    \frac{1}{g_q}\frac{\beta}{\sqrt{kT_\mathrm{e}}} S_{pq} \exp\left[
        -\frac{\omega_{pq}}{kT_\mathrm{e}}
    \right],
\end{equation}
where $k$ is the Boltzmann constant.
Although many of \textit{ab initio} calculation codes utilize more precise rate coefficients, several authors have adopted the above simplest approximation to derive analytical forms.
Griem~\cite{griem_1997} and Fujimoto et al.~\cite{Fujimoto} used 
\begin{equation}
    \label{eq:rate_fujimoto}
    \beta = \frac{2^{5/2}}{3} \pi^{1/2} \alpha a_0^2 c \sqrt{E_\mathrm{H}} \frac{1}{e^2 a_0^2}
\end{equation}
to reduce the LTE criterion for hydrogen-like ions.
Mewe has proposed a slightly different form~\cite{Mewe1972},
\begin{equation}
    \label{eq:rate_mewe}
    \beta = \frac{2^{7/2}}{3\sqrt{3}} \pi^{3/2} \alpha a_0^2 c \sqrt{E_\mathrm{H}} \frac{1}{e^2 a_0^2} \xi,
\end{equation}
where $\xi$ is the gaunt factor, which is $\xi \approx 0.15$ for transitions associated with a change in the principal quantum number (which results in almost the same value to \eref{eq:rate_fujimoto}), and $\xi \approx 0.6$ for those without the change.

Once excitation rate coefficients are calculated, either from accurate methods or the easy methods, the corresponding deexcitation rates coefficients $F_{p\gets q}$ can be deduced from the detailed balance principle,
\begin{equation}
    \label{eq:detailed_balance}
    F_{p\gets q} = \frac{g_p}{g_q}C_{q \gets p} \exp\left[\frac{\omega_{pq}}{k T_\mathrm{e}}\right].
\end{equation}

\subsection{Flexible Atomic Code\label{sec:manyelectron_crmodel}}

Although an application of an \textit{ab initio} simulation code is not an essential purpose of this paper, we use it as a reference to compare with our model presented in the next Section.
Some concrete examples will also help the discussion.
Therefore, only in this subsection, we simulate the atomic structure, rate coefficients, and the resultant population distribution for some example atoms with an \textit{ab initio} calculation code.

There are several integrated packages of the atomic structure calculation and CR model for studying many-electron atom spectra~\cite{Cowan1981,Gu2008,amusia1997computation,Bar-Shalom2001,Hill2018}.
One common package is Flexible Atomic Code (FAC)~\cite{Gu2008}.
In the FAC, the electron wavefunction of a many-electron atom is approximated by a linear combination of single-body product wavefunctions.
Their mixing coefficients are calculated based on the configuration interaction method, and the line strengths are computed from these mixing coefficients.
The electron-impact excitation cross sections are computed via distorted wave approximation, which is believed to be more accurate than \eref{eq:excitation_rate} \cite{Gu2008}.
The electron-impact ionization cross sections and autoionization rates are also computed using the FAC (although the CCRM introduced later does not include the these processes).
Using these cross sections and rates, \eref{eq:crm} is solved, and the population distribution is simulated with given pairs of $n_\mathrm{e}$ and $T_\mathrm{e}$.

As examples of complex many-electron atoms, we consider neutral iron (Fe\textsc{I}), manganese-like iron (Fe\textsc{II}), and nickel-like krypton (Kr\textsc{IX}), which have different number of electrons and different effective charges.
The details of the configurations included in each calculation are presented in Appendix \ref{sec:fac_config}.

\begin{figure*}
    \centering
    \includegraphics[width=15.5cm]{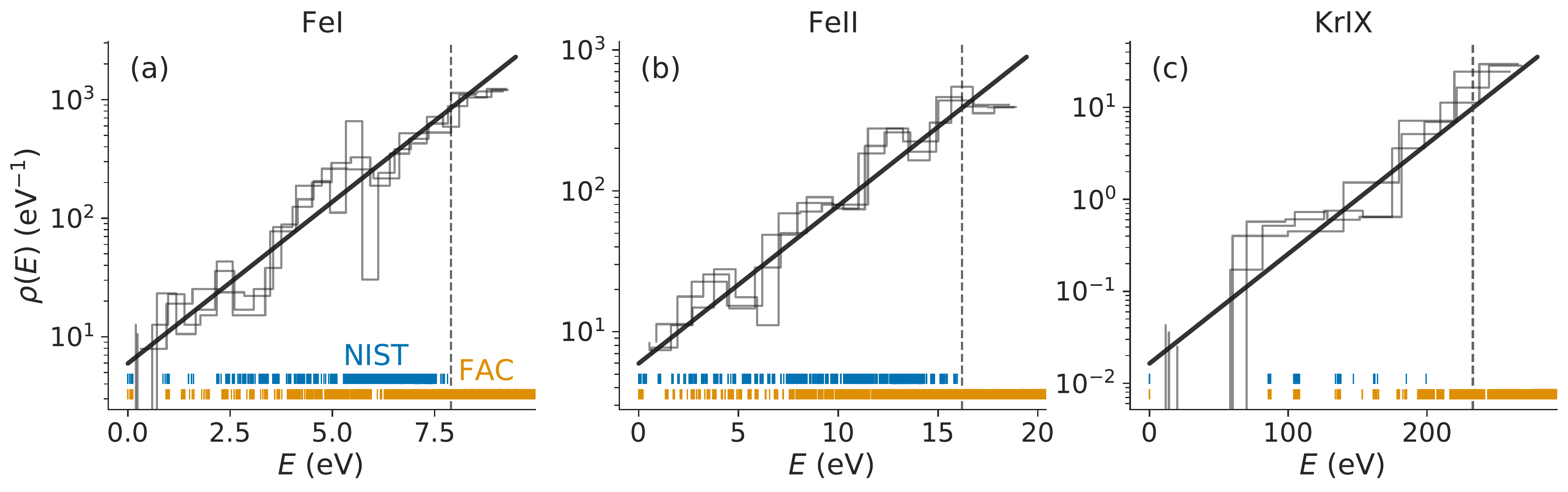}
    \caption{
        Level densities of (a) Fe\textsc{I}, (b) Fe\textsc{II} (Mn-like Fe), and (c) Kr\textsc{IX} (Ni-like krypton).
        Actual energy levels compiled in NIST ASD and computed using the FAC are shown by vertical bars in each figure.
        The gray step lines are histograms for the computed energy levels with three different bin sizes.
        The black solid lines are fit by the constant-temperature model (\eref{eq:level_density}).
        The vertical dashed lines indicate the first ionization energy.
    }
    \label{fig:level_density}
\end{figure*}

The vertical bars in Fig.~\ref{fig:level_density} (a), (b), and (c) show the energy levels calculated with FAC.
For comparison, the energy levels compiled in Atomic Spectra Database by National Institute of Standards and Technology (NIST ASD~\cite{NIST_ASD}) are also shown in Fig.~\ref{fig:level_density} by vertical bars.

\begin{figure*}
    \centering
    \includegraphics[width=15.5cm]{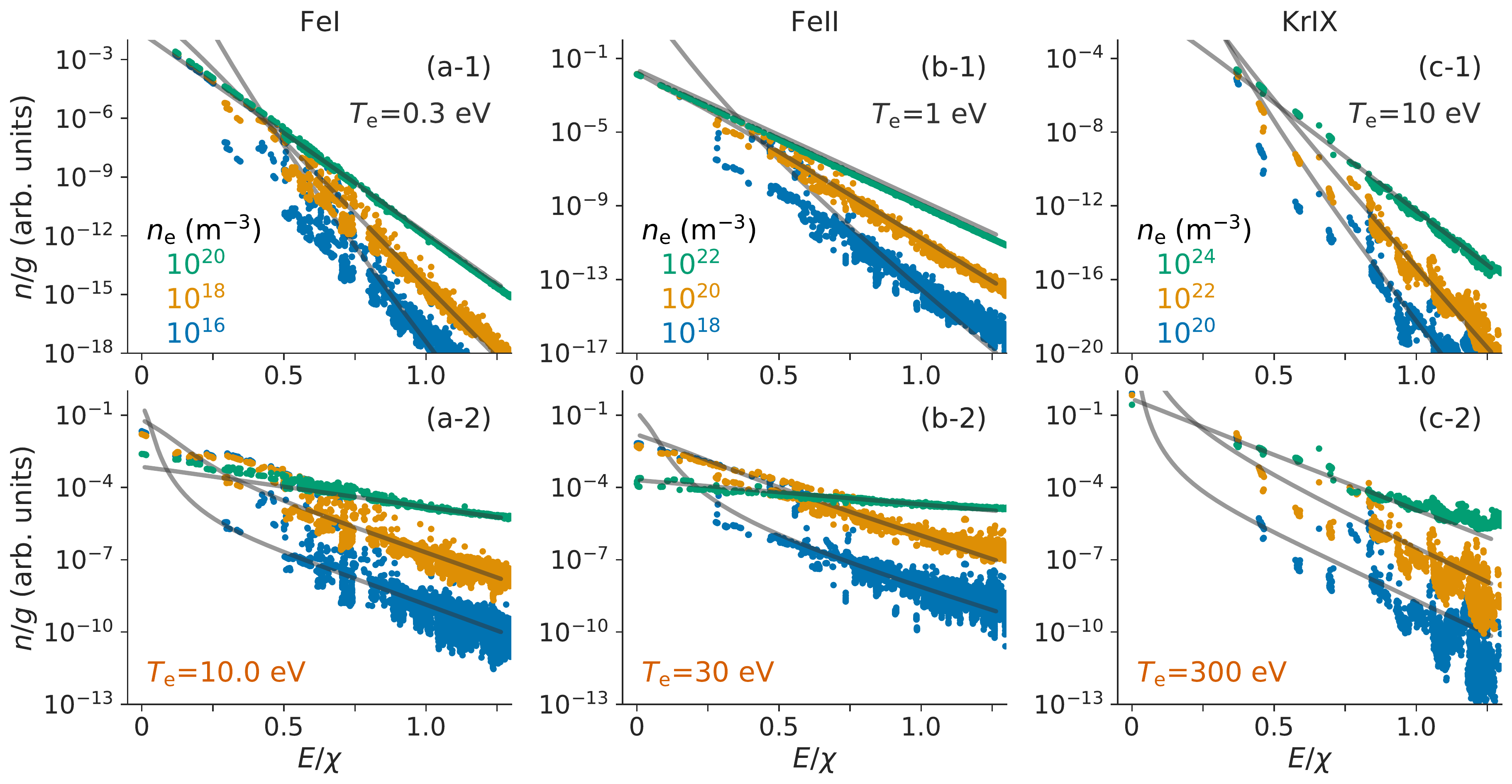}
    \caption{
        Population distribution $n/g$ of (a) Fe\textsc{I}, (b) Fe\textsc{II}, and (c) Kr\textsc{IX} computed using the FAC.
        The upper panels show the computation with $T_\mathrm{e}$ values smaller than $\epsilon_0$, and the lower panels show those with larger values.
        Three different colored dots in each panel show the results with different $n_\mathrm{e}$ values.
        In higher density, the population distribution stays on one line, \eref{eq:boltzmann}.
        At lower density, the population is scattered around an exponentially decaying line (a straight line in the semilogarithmic plot) with a steeper slope.
        The gray solid lines in each panel show the result of our simplified model with Eqs. (\ref{eq:Ain})--(\ref{eq:Fout}) with the same $T_\mathrm{e}$ and $n_\mathrm{e}$ values.
        Note that the scales of these lines are adjusted to fit those of the FAC result, and therefore, only the slope is important.
    }
    \label{fig:population}
\end{figure*}

The population distributions normalized by the statistical weight with several $T_\mathrm{e}$ and $n_\mathrm{e}$ combinations are shown in Fig.~\ref{fig:population}.
In high-electron density plasmas, the distributions become closer to the Boltzmann distribution
\begin{equation}
    \label{eq:boltzmann}
    n_p / g_p \propto \exp\left(-\frac{E_p}{kT_\mathrm{e}}\right),
\end{equation}
where $E_p$ is the excited energy of state $p$.
In low-electron density plasmas, the distribution deviates from \eref{eq:boltzmann}, but it still decreases exponentially against the excited energy particularly in highly excited states.
In lower density plasma, the slope becomes steeper and the scatter becomes bigger.
This observation is consistent with the previous works \cite{Blancard2003,Li2010,Hansen2006,Bauchea}.

\section{Continuous CR Model for Many-Electron Atoms\label{sec:model}}

In this work, we present an analytic form of \eref{eq:crm}, using the statistical theory of the atomic structure of many-electron atoms.
In particular, we assume the following two properties;
\begin{enumerate}
    \item exponential increase in the level density over the excited energy, and
    \item independently and identically distributed line strength.
\end{enumerate}
These probabilistic assumptions, as well as the huge number of energy levels and transitions, allow us to approximate the summations in \eref{eq:crm_each_term} as integrals.

In subsection \ref{subsec:atomic_structure}, we present the details of the probabilistic assumptions.
In subsection \ref{subsec:continous_crm}, we construct the CCRM by approximating the summations in \eref{eq:crm_each_term} by integrals.
In subsection \ref{subsec:continous_crm_highly}, we focus on highly excited states, which allows us to further simplify the CCRM.

\subsection{Atomic Structure Approximation\label{subsec:atomic_structure}}
\subsubsection{Level density of many-electron atoms}

It is known that in fermionic many-body systems, the level density $\rho(E)$, the number of excited states per unit energy, has a nearly exponential dependence on the excited energy.
Step lines in Fig.~\ref{fig:level_density} (a), (b), and (c) show the level densities of Fe\textsc{I}, Fe\textsc{II}, and Kr\textsc{IX}, respectively, which are computed from the simulated energy levels using the FAC.
In order to present an overall dependence without the finite bin-size effect, we show three histograms with different bin sizes for each atom.
It is clear that the level density of these many-electron atoms increases nearly exponentially.

Two most common models of the level densities of another fermionic many-body system, heavy nuclei, are back-shifted fermi gas model and constant-temperature model~\cite{TerHaar1949, VonEgidy1986}.  
Applications of these models to the level densities of many-electron atoms have been also reported by several authors~\cite{Flambaum1994,Dzuba2010}.
In the constant-temperature model, the level density is expressed as follows:
\begin{equation}
    \label{eq:level_density}
    \rho(E) = \rho_0 \exp\left(\frac{E}{\epsilon_0}\right),
\end{equation}
where $\epsilon_0$ is an energy scale parameter indicating the inverse of the level density increase rate. This value in principle can be estimated based on the number of valence electrons and shell-separation energy~\cite{Dzuba2010}.

The solid lines in Fig.~\ref{fig:level_density} show the fit using the constant-temperature model, where $\rho_0$ and $\epsilon_0$ are adjusted so that \eref{eq:level_density} matches the computed histogram.
The histograms shown in the figure are in good agreement with \eref{eq:level_density}, particularly in the highly excited energy regions.

\begin{figure}
    \centering
    \includegraphics[width=7cm]{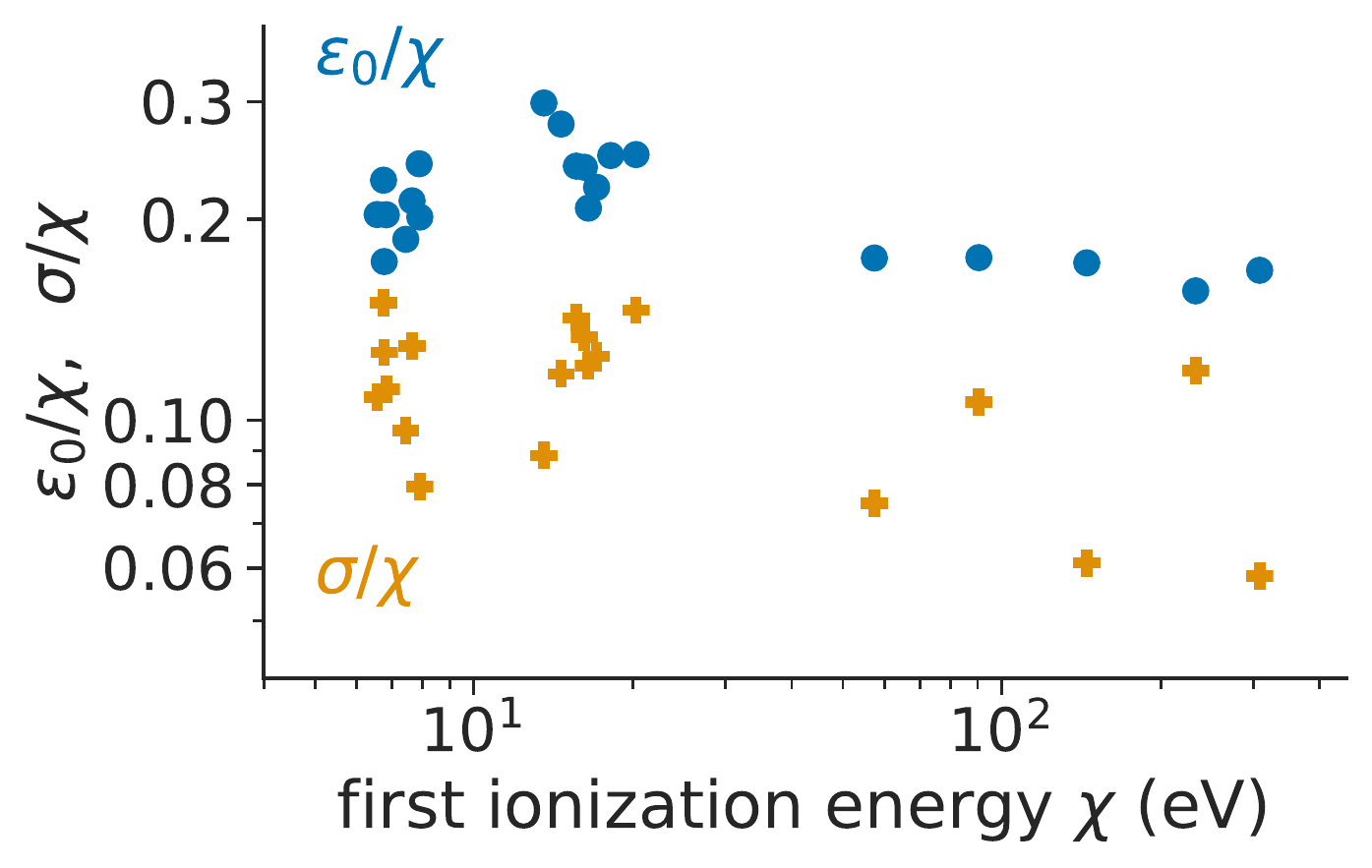}
    \caption{
        The scale factors of the level density ($\epsilon_0$) and line strength ($\sigma$) for several atomic ions.
        The values are normalized by their first ionization energy $\chi$.
        Data points around $\chi \approx 8$ eV are for neutral transition metals from Sc to Ni, and 
        those around $\chi \approx 15$ eV are for singly charged ions of transition metals.
        The other five points are the isoelectronic sequence of Fe and Ni.
    }
    \label{fig:epsilon_sigma}
\end{figure}

Figure \ref{fig:epsilon_sigma} shows the ionization-energy dependence of the scale parameter $\epsilon_0$ for several atomic ions.
In the figure, the computed values of $\epsilon_0$ for neutral and singly charged ions of transition metals and the isoelectronic sequence of Fe-like and Ni-like ions are shown.
The first ionization energy data $\chi$ are taken from NIST ASD~\cite{NIST_ASD}.
$\epsilon_0 / \chi$ is almost constant over the wide variety of atoms and charges.
\begin{equation}
    \label{eq:epsilon_scaling}
    \epsilon_0 / \chi \approx 0.2
\end{equation}
may be a good empirical approximation for transition metals.
Atomic parameters for the atoms used in this work are presented in Table~\ref{tab:atomicparams}.
Note that the uncertainty of $\epsilon_0$ is $\approx$ 20\%, which originates from the variation of the FAC computation result over the change in hyper-parameters, e.g. the number of basis states and central potential.

\begin{table}[htb]
    \centering
    \caption{Scale factors of the level structure $\epsilon_0$~\eref{eq:level_density} and line strength  $\sigma$~\eref{eq:strength-function} used in this work. The ionization energy $\chi$ is also shown.}
    \label{tab:atomicparams}
    \begin{ruledtabular}
    \begin{tabular}{lccc}
    Ion & $\epsilon_0$ (eV) & $\sigma$ (eV) & $\chi$ (eV) \\
    \hline
    FeI & 1.6 & 0.62 & 7.902 \\
    FeII & 3.9 & 2.3 & 16.2 \\
    KrIX & 36 & 23 & 233.0 \\
    \end{tabular}
    \end{ruledtabular}
\end{table}

\subsubsection{Line strength distribution}

The line strength distribution of fermionic many-body systems has been approximated as independent and identical \cite{Bohr1936}.
This approximation is valid if the wavefunction mixing is so large that the system can be viewed as a quantum chaotic system.
For heavy nuclei, the line strength distribution was approximated by the Porter--Thomas distribution~\cite{Porter1956, Grimes1983, Bisson1991},
\begin{equation}
    \label{eq:porter-thomas}
    p(S|\omega) = 
    \frac{1}{\sqrt{2 \pi S \overline{S}(\omega)}} \exp\left(-\frac{S}{2\overline{S}(\omega)}\right),
\end{equation}
where $\overline{S}(\omega)$ is the mean value of $S$, the so-called gamma strength function, which is only a function of the energy difference $\omega$ of the initial and final states.
This distribution was also tested for the line strength distribution of many-electron atoms~\cite{flambaum94pra, Flambaum1998}
Equation (\ref{eq:porter-thomas}) is based on the Brink--Axel hypothesis~\cite{Brink1955, Axel1962}, where the line strength depends only on the transition energy, but does not depend on properties of the initial and final states.
Recently, experimental evidence of this hypothesis for heavy nuclei has been reported~\cite{Guttormsen2016}.

We also adopt this assumption, as in Ref.~\cite{Fujii2020}, for many-electron atoms.
In Ref.~\cite{Fujii2020}, we assumed a uniform line strength distribution over the energy difference,
$\overline{S}(\omega) \propto \omega^{0}$, for simplicity.
However, in this work, we adopt a more realistic assumption for $\overline{S}(\omega)$ that can be consistent with the sum rule of the oscillator strength,
\begin{align}
    \notag
    N &= \sum_j f_{i \to j} \\
    \notag
    & \approx 
    \int_0^E f(E, E-\omega) \rho(E-\omega)\mathrm{d}\omega\\
    \label{eq:sum_rule}
    & + \int_0^\infty f(E, E+\omega) \rho(E+\omega)\mathrm{d}\omega,
\end{align}
where $N$ is the number of electrons in the atom, and $f(E, E + \omega)$ is the oscillator strength from the state at $E$ to that at $E+\omega$.
With $ f \propto S \omega \propto \omega^1$, the second integral in the right hand side of \eref{eq:sum_rule} diverges.
Although there are no well-established model distributions for line strengths of many-electron atoms, we assume an exponentially decreasing function,
\begin{equation}
    \label{eq:strength-function}
    \overline{S}(\omega) = S_0\exp\left(-\frac{|\omega|}{\sigma}\right)
\end{equation}
where $\sigma$ is a scale parameter.
With $\sigma < \epsilon_0$, \eref{eq:sum_rule} does not diverge. \eref{eq:strength-function} can be understood from the statistical theory because the line strength between orbitals near $E$ and orbitals near $E+\omega$ is distributed over the ergodically mixed states, resulting in an exponential decay.

\begin{figure*}
    \centering
    \includegraphics[width=15.5cm]{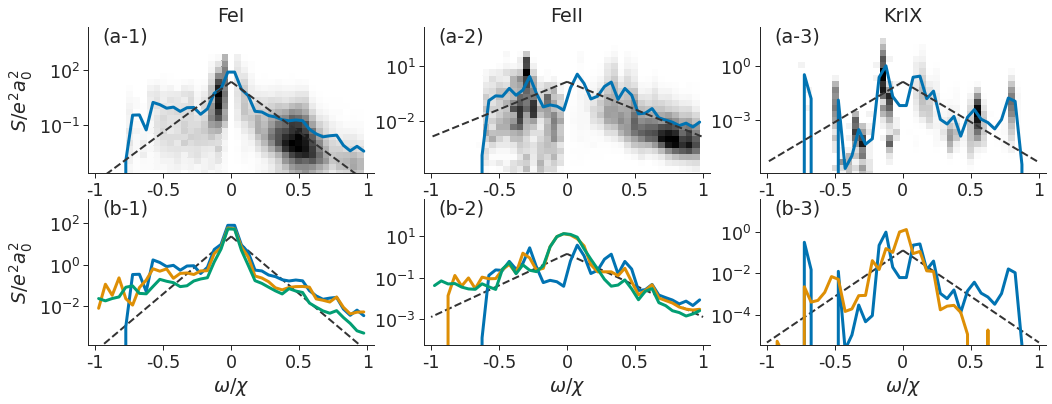}
    \caption{
        Strength functions for the line strengths of Fe\textsc{I}, Fe\textsc{II}, and Kr\textsc{IX}.
        (a) 
        The density distribution of the line strength values as a function of the transition energy $\omega$ and their actual values. 
        Note that the density scale is different in $\omega < 0$ and $\omega > 0$. 
        The solid lines indicate their averaged values, i.e., the strength function $\overline{S}(\omega)$, computed from these data.
        The dotted slope lines indicate \eref{eq:strength-function}, the energy scale $\sigma$ of which is computed from the weighted expectation.
        (b) 
        Initial energy dependence of $\overline{S}(\omega)$.
        4--6, 6--8, and 8--10 eV for Fe\textsc{I}; 6--10, 10--14, and 14--18 eV for Fe\textsc{II}; and 140--200 and 200--240 eV for Kr\textsc{IX} are shown.
    }
    \label{fig:strength_function}
\end{figure*}

Figure~\ref{fig:strength_function} (a) shows the density distribution of $S$ values computed using the FAC for Fe\textsc{I}, Fe\textsc{II}, and Kr\textsc{IX}, as a function of the energy difference $\omega$ and its actual value.
The strength function values are computed by averaging these points in certain energy bins, and are shown by solid lines.
This function decreases exponentially against the transition energy.
Our modeled strength functions \eref{eq:strength-function} are shown by dotted lines in the figure, 
$\sigma$ values of which are estimated from the computed $S$ values.
In Fig.~\ref{fig:strength_function} (b), the strength functions computed from the different energy range of the initial states are shown.
Although a slight initial-energy dependence can be seen, this approximation agrees well with the FAC computation.

In Fig.~\ref{fig:epsilon_sigma}, we also plot $\sigma$ values for several transition metals and their isoelectronic sequence.
Although it shows more scatter than that of $\epsilon_0$, 
\begin{equation}
    \label{eq:sigma_scaling}
    \sigma / \chi \approx 0.1,
\end{equation}
may be a reasonable estimate.

\subsection{Continuous Balance Equation\label{subsec:continous_crm}}

The spontaneous transition rate can be directly computed from \eref{eq:radiative_rate} with given $S$ distributions.
Let us approximate $\mathcal{A}_p^\mathrm{in}$ by using \eref{eq:level_density} and \eref{eq:radiative_rate}.
\begin{align}
    \notag
    \mathcal{A}_p^\mathrm{in} &= \sum_{q>p} A_{p\gets q} n_q \\
    \notag
    & \approx \int_0^\infty 
        \gamma \omega^3 \overline{S}(\omega) \rho(E_p + \omega) n(E_p + \omega) \mathrm{d}\omega 
    \\
    \label{eq:Ain}
    & = \gamma \rho_0 S_0 e^{E_p/\epsilon_0}
        \int_0^\infty 
        \omega^3 e^{
            -\omega / \delta
        }
        n(E_p + \omega) \mathrm{d}\omega 
\end{align}
with 
\begin{equation}
    \delta = \frac{1}{\frac{1}{\sigma} - \frac{1}{\epsilon_0}}.
\end{equation}
$n(E_p)$ is the population of the state $p$.
Here, we approximate $g \approx \overline{g}$ and $S \approx \overline{S}(\omega)$, where $\overline{g}$ is the mean value of the statistical weight.
Note that this approximation comes from the central limit theorem and is valid if the distribution of $S$ is \textit{independent}.
Therefore, the shape of the distribution is not restricted to \eref{eq:porter-thomas}, but is rather arbitrary.

Similarly, the total radiative outflux $\mathcal{A}_p^\mathrm{out}$ can be approximated as
\begin{widetext}
\begin{align}
    \notag
    \mathcal{A}_p^\mathrm{out} &= \sum_{q<p} A_{q\gets p} n_p 
    \approx \int_0^{E_p} 
        \gamma \omega^3 \overline{S}(\omega) \rho(E_p - \omega) n(E_p) \mathrm{d}\omega \\
    \label{eq:Aout}
    & = \gamma \rho_0 S_0 e^{E_p/\epsilon_0}
    \left[
    6 \mu^4 - 
    \left(6\mu^4 + 6\mu^3 E_p + 3\mu^2 E_p^2 + \mu E_p^3\right) e^{- E_p / \mu}
    \right]
    n(E),
\end{align}
with
\begin{equation}
    \mu = \frac{1}{\frac{1}{\sigma} + \frac{1}{\epsilon_0}}.
\end{equation}
Using \eref{eq:epsilon_scaling} and \eref{eq:sigma_scaling} we may further approximate $\delta \approx \chi/5$ and $\mu \approx \chi/15$.

Although the electron impact excitation rate has a much more complex dependence on $S$, we adopt the simpler approximation \eref{eq:rate_mewe} with $\xi = 0.6$~\cite{Mewe1972}.
With this dependence, we can also approximate $\mathcal{C}^\mathrm{in}$, $\mathcal{C}^\mathrm{out}$, $\mathcal{F}^\mathrm{in}$, and $\mathcal{F}^\mathrm{out}$ as follows:
\begin{align}
    \notag
    \mathcal{C}_p^\mathrm{in} &= \sum_{q<p} C_{p\gets q}(T_\mathrm{e}) n_\mathrm{e} n_\mathrm{q}
    \approx \int_0^{E_p} \overline{C}(T_\mathrm{e}, \omega) n_\mathrm{e} n(E_p-\omega) \rho(E_p-\omega) \mathrm{d}\omega\\
    & \label{eq:Cin}
    =
    \beta \frac{n_{\mathrm{e}}}{\sqrt{kT_\mathrm{e}}} \rho_0 S_0 e^{E_p/\epsilon_0}
    \int_0^{E_p} 
    \exp \left[
    -\left( \frac{1}{kT_{\mathrm{e}}} + \frac{1}{\mu} \right) \omega
    \right] n(E_p - \omega) \mathrm{d}\omega
    \\
    %
    \notag
    \mathcal{C}_p^\mathrm{out} &= \sum_{q>p} C_{q\gets p}(T_\mathrm{e}) n_\mathrm{e} n_\mathrm{p}
    \approx \int_0^\infty \overline{C}(T_\mathrm{e}, \omega) n_\mathrm{e} n(E_p) \rho(E_p+\omega) \mathrm{d}\omega\\
    &  \label{eq:Cout}
    =
    \beta \frac{n_{\mathrm{e}}}{\sqrt{kT_\mathrm{e}}} \rho_0 S_0 e^{E_p/\epsilon_0}
    \frac{1}{\frac{1}{k T_\mathrm{e}}+\frac{1}{\delta}} n(E_p)
    \\
    %
    \notag
    \mathcal{F}_p^\mathrm{in} &= \sum_{q>p} F_{p\gets q}(T_\mathrm{e}) n_\mathrm{e} n_\mathrm{q}
    \approx \int_0^\infty \overline{F}(T_\mathrm{e}, \omega) n_\mathrm{e} n(E_p+\omega) \rho(E_p+\omega) \mathrm{d}\omega\\
    & \label{eq:Fin} 
    =
    \beta \frac{n_{\mathrm{e}}}{\sqrt{kT_\mathrm{e}}} \rho_0 S_0 e^{E_p/\epsilon_0}
    \int_0^\infty
    e^{-\omega / \delta}
    n(E_p + \omega) \mathrm{d}\omega
    \\
    %
    \notag
    \mathcal{F}_p^\mathrm{out} &= \sum_{q<p} F_{q\gets p}(T_\mathrm{e}) n_\mathrm{e} n_\mathrm{p}
    \approx \int_0^{E_p} \overline{F}(T_\mathrm{e}, \omega) n_\mathrm{e} n(E_p) \rho(E_p-\omega) \mathrm{d}\omega\\
    & \label{eq:Fout}
    =
    \beta \frac{n_{\mathrm{e}}}{\sqrt{kT_\mathrm{e}}} \rho_0 S_0 e^{E_p/\epsilon_0}
    \mu\left(1-e^{-E_p / \mu}\right) n(E_p).
\end{align}
\end{widetext}
Here, we again use the averaged value of $S_{pq} \approx \overline{S}(\omega_{pq}) = S_0 \exp(-|\omega_{pq}| / \sigma)$.
$\overline{C}$ and $\overline{F}$ are the rate coefficients defined in \eref{eq:excitation_rate} and \eref{eq:detailed_balance}, respectively, with $S$ substituted by $\overline{S}$.

We solve \eref{eq:crm} by substituting Eqs. (\ref{eq:Ain} -- \ref{eq:Fout}).
Solid lines in Fig.~\ref{fig:population} show the numerical solutions, whose vertical scales ($n_0$) are chosen to fit the FAC results.
Although our CCRM assumes a smooth population distribution and therefore does not reproduce the vertical scatter (relative scatter up to $10^2$ depending on $n_\mathrm{e}$ and $T_\mathrm{e}$), its slope agrees well with the FAC results, particularly in highly excited states.
Recall that in our model, we only use two parameters to represent the atomic structure, $\epsilon_0$ and $\sigma$.
In light of this huge simplification, the consistency between our CCRM and the FAC computation is surprising.
This suggests that the probabilistic approximation is reasonable for many-electron atoms.

\subsection{Further simplification for highly excited states\label{subsec:continous_crm_highly}}

In highly excited states with $E \gg \epsilon_0$, the following approximation for the finite-range integration in \eref{eq:Ain}, \eref{eq:Cin}, and \eref{eq:Fin} may be valid,
\begin{align}
    \label{eq:infinte_approximation}
    \int_0^E \cdot\; \mathrm{d}\omega \approx \int_0^\infty \cdot\; \mathrm{d}\omega, 
\end{align}
as the integrands decreases quickly for large $\omega$ due to the exponential decrease in the rate coefficient over the energy difference $\omega$.
Note that ``$\cdot$'' indicates any of the integrands in \eref{eq:Ain}, \eref{eq:Cin}, and \eref{eq:Fin}.
Then, all Eqs. (\ref{eq:Ain}) -- (\ref{eq:Fout}) become translation-invariant, i.e., $n(E) \propto n(E + \Delta)$ holds for any energy $E$.
Therefore, $n(E)$ should have the following form
\begin{equation}
    \label{eq:asymptotic}
    n(E) = n_0 \exp\left(-\frac{E}{k T_\mathrm{ex}}\right),
\end{equation}
with an effective excitation temperature $T_\mathrm{ex}$ and an arbitrary scale $n_0$.

By substituting \eref{eq:asymptotic} into Eqs. (\ref{eq:Ain}) -- (\ref{eq:Fout}) and assuming $E \gg \epsilon_0$, we get
\begin{align}
    \label{eq:Ain_highE}
    \mathcal{A}^\mathrm{in} 
    \approx &
    \gamma \rho_0 S_0 n_0 e^{-E/\tau}
    \frac{6}{\left(\frac{1}{k T_\mathrm{ex}} + \frac{1}{\delta}\right)^4}
    \\
    \label{eq:Aout_highE} 
    \mathcal{A}^\mathrm{out} 
    \approx &
    \gamma \rho_0 S_0 n_0 e^{-E/\tau}
    6 \mu^4
    \\
    \label{eq:Cin_highE} 
    \mathcal{C}^\mathrm{in} 
    \approx &
    \beta n_\mathrm{e}\sqrt{k T_\mathrm{e}}
    \rho_0 S_0 n_0 e^{-E/\tau}
    \frac{1}{1 + \frac{kT_\mathrm{e}}{\mu} - \frac{kT_\mathrm{e}}{kT_\mathrm{ex}}}
    \\
    \label{eq:Cout_highE} 
    \mathcal{C}^\mathrm{out} 
    \approx &
    \beta n_\mathrm{e}\sqrt{k T_\mathrm{e}}
    \rho_0 S_0 n_0 e^{-E/\tau}
    \frac{1}{1 + \frac{k T_\mathrm{e}}{\delta}}
    \\
    \label{eq:Fin_highE} 
    \mathcal{F}^\mathrm{in} 
    \approx &
    \beta n_\mathrm{e}\sqrt{k T_\mathrm{e}}
    \rho_0 S_0 n_0 e^{-E/\tau}
    \frac{1}{\frac{k T_\mathrm{e}}{\delta} + 
             \frac{k T_\mathrm{e}}{k T_\mathrm{ex}}}
    \\
    \label{eq:Fout_highE} 
    \mathcal{F}^\mathrm{out} 
    \approx &
    \beta n_\mathrm{e}\sqrt{k T_\mathrm{e}}
    \rho_0 S_0 n_0 e^{-E/\tau}
    \frac{\mu}{kT_\mathrm{e}},
\end{align}
with 
\begin{align}
    \tau &= \frac{1}{\frac{1}{kT_\mathrm{ex}} - \frac{1}{\epsilon_0}}
\end{align}
By substituting Eqs. (\ref{eq:Ain_highE})--(\ref{eq:Fout_highE}) into \eref{eq:crm}, we have the following equation for $T_\mathrm{ex}$,
\begin{align}
    \notag
    6\gamma \left\{
        \frac{1}{\left(\frac{1}{k T_\mathrm{ex}} + \frac{1}{\delta}\right)^4} 
        - \mu^4
    \right\}
    + \beta n_\mathrm{e} \sqrt{k T_\mathrm{e}} \Biggl\{
    \frac{1}{1 + \frac{kT_\mathrm{e}}{\mu} - \frac{kT_\mathrm{e}}{kT_\mathrm{ex}}}
    \\
    \label{eq:final}
    - \frac{1}{1 + \frac{k T_\mathrm{e}}{\delta}}
    + \frac{1}{\frac{k T_\mathrm{e}}{\delta} + 
            \frac{k T_\mathrm{e}}{k T_\mathrm{ex}}}
    - \frac{\mu}{k T_\mathrm{e}}
    \Biggr\} = 0.
\end{align}
Although \eref{eq:final} is not analytically solvable, its numerical solution can be found easily with given $T_\mathrm{e}, n_\mathrm{e}, \epsilon_0$, and $\sigma$.

\begin{figure*}
    \centering
    \includegraphics[width=15.5cm]{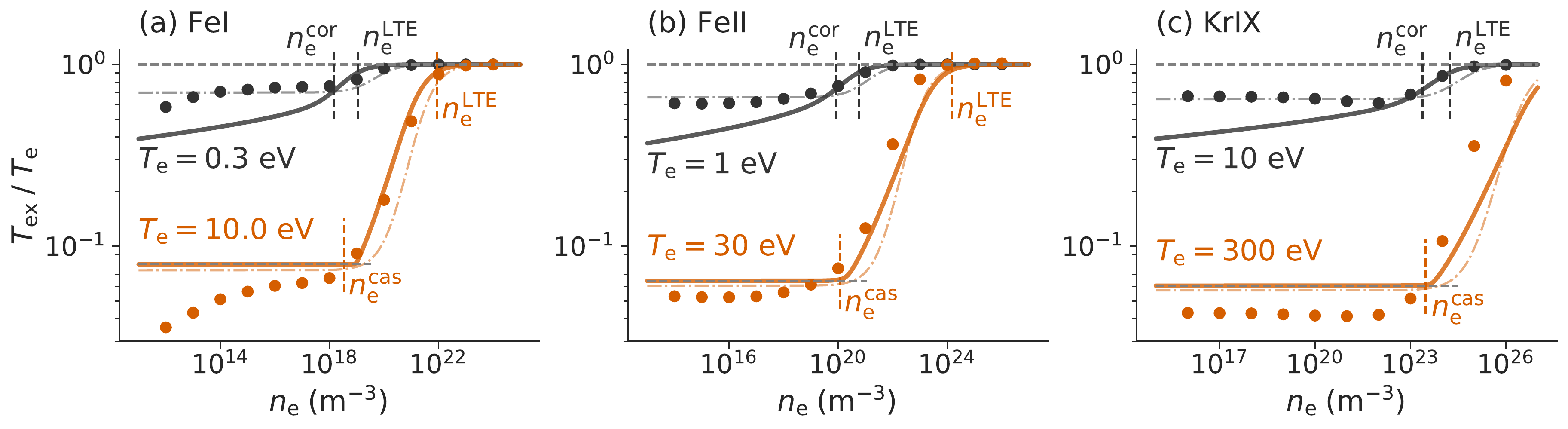}
    \caption{
        $n_\mathrm{e}$ dependence of $T_\mathrm{ex}$ for (a) Fe\textsc{I}, (b) Fe\textsc{II}, and (c) Kr\textsc{IX}.
        Markers are generated from the FAC results, and lines are computed from \eref{eq:final}, at $E \approx 0.7 \chi$.
        The two horizontal dotted lines in each panel show $T_\mathrm{ex} = T_\mathrm{e}$ and $T_\mathrm{ex} = \epsilon_0 / 2k$.
        The four vertical dotted lines indicate the boundary densities, 
        \eref{eq:ne_coronal}, \eref{eq:ne_sat_lowT}, \eref{eq:ne_cascade}, and \eref{eq:ne_sat_highT}, respectively.
        The chain curves show the $T_\mathrm{ex}$ values estimated from two-level model \eref{eq:Teff_2lev} with $\Delta E = \frac{3}{2}\epsilon_0$.
    }
    \label{fig:Tex}
\end{figure*}

In Fig.~\ref{fig:Tex}, we present the excitation temperatures for the FAC results and the CCRM.
In order to estimate $T_\mathrm{ex}$ from the result obtained using FAC, we choose a certain excited energy range and fit the result in this region by \eref{eq:asymptotic} (see caption of Fig.~\ref{fig:Tex} for the details).
In the high-density limit, $T_\mathrm{ex}$ approaches $T_\mathrm{e}$.
On the other hand, in the low-density limit, $T_\mathrm{ex}$ approaches a different value.
There is a density region where the transition between these two phases takes place.
These tendencies, and the actual values of $T_\mathrm{ex}$, are consistent between FAC and our CCRM.

\section{Discussions\label{sec:discussions}}

In Fig.~\ref{fig:population}, we see that the excited state population distribution changes depending on $n_\mathrm{e}$ and $T_\mathrm{e}$ values of the plasma.
In this section, we aim to \textit{understand} this population kinetics using our CCRM.

The discussion in this section is largely inspired by the proceeding works for H-like ions by Fujimoto~\cite{Fujimoto}.
They summarized the population kinetics of H-like ions and its $n_\mathrm{e}$ and $T_\mathrm{e}$ dependence.
Figure~\ref{fig:diagram}~(a) shows the diagram of population kinetics of H-like ions.
There are two typical density regions, corona phase and LTE phase (or saturation phase in \cite{Fujimoto, Fujimoto1990}), in which the population kinetics is systematically different.
In the corona phase, excited state atoms are dominantly generated by electron-impact excitation from the ground state, while being dominantly depopulated by radiative decay.
On the other hand, in higher density plasmas, the dominant population path is the excitation from the next lower level and the dominant depopulation path is the excitation to the next higher level.
The density boundary is given by~\cite{Griem,Fujimoto}
\begin{align}
    \label{eq:boundary_hydrogen}
    n_\mathrm{e} \approx 0.7\cdot 3\cdot 2^3 \frac{\gamma}{\beta} \left(\frac{z^2 E_\mathrm{H}}{2}\right)^3 (kT_\mathrm{e})^{1/2} \nu^{-8.5},
\end{align}
where $\nu$ represents the principal quantum number of the level and $z^2 E_\mathrm{H} / 2$ corresponds to the ionization energy of an H-like ion with nuclear charge $z$.
More details can be found in Appendix~\ref{sec:hydrogen}.

We will discuss the population kinetics of many-electron atoms.
As we will see below, their population kinetics is also systematically different depending on the $n_\mathrm{e}$ and $T_\mathrm{e}$ values in plasmas.
Although the population distribution of many-electron atoms is found to be Boltzmann-like \eref{eq:asymptotic}, which is in contrast with the power-law distribution of H-like ions (see \eref{eq:minus_half} and \eref{eq:minus_sixth} in Appendix), the density boundaries \eref{eq:ne_coronal} and \eref{eq:ne_cascade} show a similar form to that for H-like ions \eref{eq:boundary_hydrogen}.

\subsection{Highly Excited States}

\begin{figure*}
    \centering
    \includegraphics[width=12.5cm]{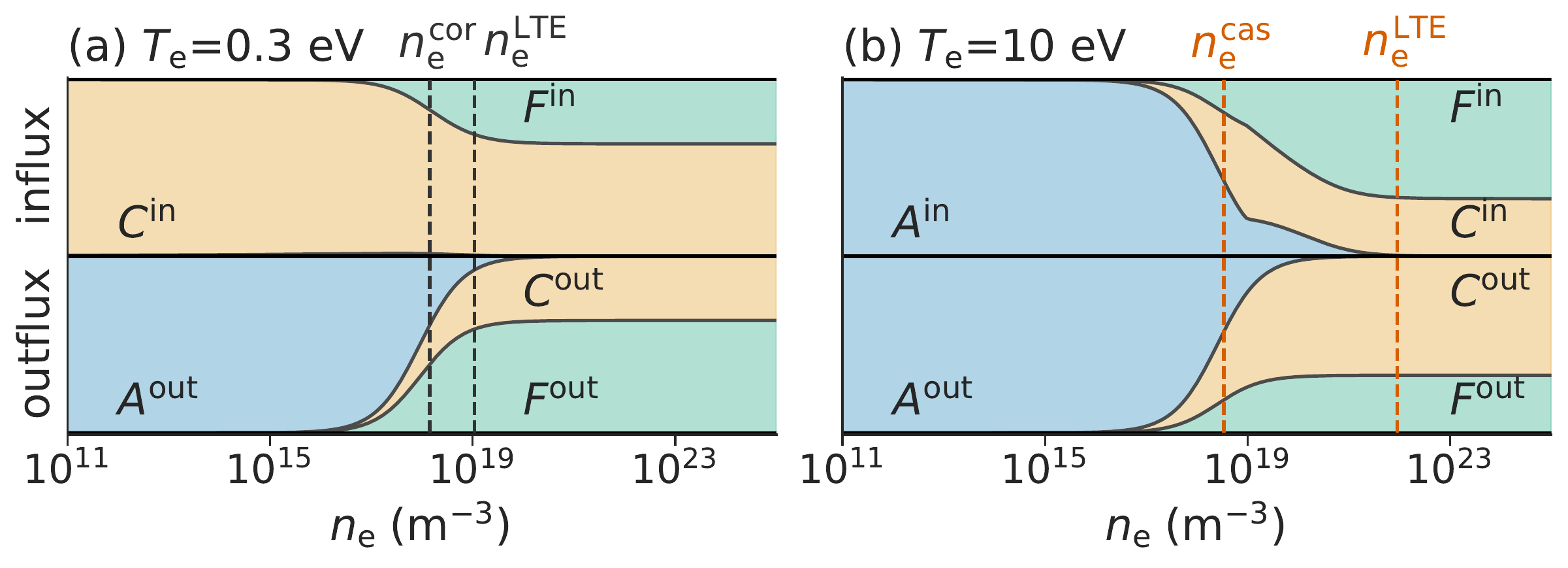}
    \caption{
        Flux composition for Fe\textsc{I} at (a) $T_\mathrm{e} = 0.3$ eV and (b) $T_\mathrm{e} = 10$ eV. Here $\mathcal{A}$ is due to spontaneous emission processes, $\mathcal{C}$ is collisional excitation, and $\mathcal{F}$ is collisional de-excitation.
    }
    \label{fig:fluxes}
\end{figure*}

\subsubsection{Low-temperature plasmas}

In Fig.~\ref{fig:fluxes}~(a), we show the flux composition for highly excited states of Fe\textsc{I} with $T_\mathrm{e} = 0.3$ eV.
The upper part of the figure shows the influx to a certain level, and the bottom part is the outflux from this level.
The total influx and outflux are normalized to unity, so that their compositions can be clearly seen.

On the low-$n_\mathrm{e}$ side, the dominant influx is collisional excitation from lower levels, and the dominant outflux is a spontaneous decay to lower levels.
On the high-$n_\mathrm{e}$ side, the dominant influx does not change, collisional excitation, and the dominant outflux becomes collisional de-excitation to lower levels.

\paragraph{Low-density region: corona phase ---}

The low-$n_\mathrm{e}$ region is similar to the corona phase, which is defined for H-like ions.
In the original corona phase, the dominant influx is collisional excitation from the ground state, whereas in our case, the excitation from all lower levels is considered.

\paragraph{High-density region: LTE phase --}

In the high-$n_\mathrm{e}$ region, the effect of the spontaneous decay is negligible, and 
$\mathcal{C}^\mathrm{in} = \mathcal{F}^\mathrm{out}$ and $\mathcal{F}^\mathrm{in} = \mathcal{C}^\mathrm{out}$ are satisfied based on the principle of detailed balance.
We call this phase the LTE phase, similar to that in H-like ions.
The population influx and outflux with lower levels are dominant compared with those with higher levels.
This can be understood by considering the asymptotic value of $\mathcal{F}^\mathrm{in}$ and $\mathcal{F}^\mathrm{out}$.
If $T_\mathrm{e} \approx T_\mathrm{ex} \ll \delta$, 
\begin{align}
    \frac{\mathcal{F}^\mathrm{in}}{\mathcal{F}^\mathrm{out}} 
    \approx 
    \frac{\mu}{k T_\mathrm{e}}
\end{align}
Therefore, if $k T_\mathrm{e} \ll \mu$, the population balance with lower levels is dominant, and if $\mu \ll k T_\mathrm{e}$, the population balance with higher levels becomes dominant (compare with Fig.~\ref{fig:fluxes} ~(b)).

\paragraph{Boundary densities ---}

Here, we define two boundary densities. The lower boundary density ($n_\mathrm{e}^\mathrm{cor}$) is at the end of the corona phase, and another one ($n_\mathrm{e}^\mathrm{LTE}$) is at the start of the LTE phase.

$n_\mathrm{e}^\mathrm{cor}$ may be defined as where $\mathcal{A}^\mathrm{out} = \mathcal{F}^\mathrm{out}$ is satisfied.
Thus, the boundary density is evaluated as
\begin{equation}
    \label{eq:ne_coronal}
    n_\mathrm{e}^\mathrm{cor} \approx \frac{6 \gamma}{\beta} \mu^3 (kT_\mathrm{e})^{1/2}.
\end{equation}
These density boundaries are shown by the vertical bars in Fig.~\ref{fig:fluxes}~(a) and Fig.~\ref{fig:Tex}.
This well reproduces the $T_\mathrm{ex}$ behavior predicted by the $\textit{ab initio}$ calculation.

If we substitute \eref{eq:epsilon_scaling} and \eref{eq:sigma_scaling} into \eref{eq:ne_coronal}, this becomes
\begin{equation}
    \label{eq:ne_coronal_scale}
    n_\mathrm{e}^\mathrm{cor} \approx 2 \times 10^{-3}\frac{\gamma}{\beta} \chi^3 (kT_\mathrm{e})^{1/2}.
\end{equation}
This form can be directly compared with the boundary density of H-like ions \eref{eq:boundary_hydrogen}.
Both boundary densities scale as $\chi^3 T_\mathrm{e}^{1/2}$.
However, the boundary for many-electron atoms does not depend on the excited energy (except for that discussed in subsection~\ref{subsec:low_excited_state}), in contrast with the $\nu^{-8.5}$-dependence in \eref{eq:boundary_hydrogen}.

We also define $n_\mathrm{e}^\mathrm{LTE}$ as the boundary density where $T_\mathrm{ex} = 0.9 T_\mathrm{e}$.
By substituting this into \eref{eq:final} we obtain
\begin{equation}
    \label{eq:ne_sat_lowT}
    n_\mathrm{e}^\mathrm{LTE} \approx \frac{60 \gamma}{\beta} \mu^2 (kT_\mathrm{e})^{3/2}.
\end{equation}
This boundary density scales as $T_\mathrm{e}^{3/2}$ in contrast with \eref{eq:ne_coronal}.
However, as can be seen in Fig.~\ref{fig:Tex} and Fig.~\ref{fig:temperature_diagram}, these two boundaries have similar values in the low-temperature region.
This is because in this region, $T_\mathrm{e} < \epsilon_0 / 2 \approx \mu$ should be satisfied, as we will see later.

\subsubsection{High-temperature plasmas}

In Fig.~\ref{fig:fluxes}~(b), we show the flux composition for highly excited states of Fe\textsc{I} with $T_\mathrm{e} = 10$ eV.
On the low-$n_\mathrm{e}$ side, the dominant influx is a spontaneous decay from higher levels, and the dominant outflux is a spontaneous decay to lower levels.
On the high-$n_\mathrm{e}$ side, the dominant influx is collisional de-excitation from higher levels, and the dominant outflux becomes collisional excitation to higher levels.

\paragraph{Low-density region: cascade phase ---}

In the low-density region, the population balance is established at $\mathcal{A}^\mathrm{in} \approx \mathcal{A}^\mathrm{out}$.
From \eref{eq:Ain_highE} and \eref{eq:Aout_highE}, $T_\mathrm{ex}$ can be reduced as follows:
\begin{equation}
    \label{eq:Tex_highT_lowN}
    T_\mathrm{ex} = \epsilon_0 / 2k.
\end{equation} 
The value of $\epsilon_0 / 2k$ is plotted in Fig.~\ref{fig:Tex} by horizontal dotted lines.
This excitation temperature agrees well with the FAC result.

\paragraph{High-density limit: LTE phase ---}

In this density region, the population balance is established at $\mathcal{F}^\mathrm{in} \approx \mathcal{F}^\mathrm{out}$.
Because the effect of spontaneous decay is negligible in this region, $T_\mathrm{ex} = T_\mathrm{e}$ is established, as in low-temperature plasmas.

\paragraph{Boundary densities ---}

We define two boundary densities for high-temperature plasmas, as in the low-temperature plasmas. The lower boundary density ($n_\mathrm{e}^\mathrm{cas}$) is at the end of the cascading phase, and another one ($n_\mathrm{e}^\mathrm{LTE}$) is at the start of the LTE phase.

As the dominant outflux changes from $\mathcal{A}^\mathrm{out}$ to $\mathcal{F}^\mathrm{out}$ when the corona phase changes to the LTE phase, $n_\mathrm{e}^\mathrm{cas}$ may be defined, where $\mathcal{A}^\mathrm{in} = \mathcal{F}^\mathrm{in}$ is satisfied.
From the equality and \eref{eq:Aout_highE} and \eref{eq:Fout_highE}, the boundary density is reduced as
\begin{equation}
    \label{eq:ne_cascade}
    n_\mathrm{e}^\mathrm{cas} \approx \frac{6 \gamma}{\beta} \mu^3 \frac{\mu}{\delta}~(kT_\mathrm{e})^{1/2}.
\end{equation}
This density boundary is shown by the vertical bars in Fig.~\ref{fig:fluxes}~(a) and Fig.~\ref{fig:Tex}.
This boundary density also scales as $\chi^3 T_\mathrm{e}^{1/2}$, similar to \eref{eq:boundary_hydrogen} and \eref{eq:ne_coronal}.

We define another boundary density $n_\mathrm{e}^\mathrm{LTE}$, where $T_\mathrm{ex} = 0.9 T_\mathrm{e}$ is satisfied.
By substituting this into \eref{eq:final},
\begin{equation}
    \label{eq:ne_sat_highT}
    n_\mathrm{e}^\mathrm{LTE} \approx \frac{60 \gamma}{\beta} \delta^2 (kT_\mathrm{e})^{3/2}.
\end{equation}
These density boundaries are shown by the vertical bars in Fig.~\ref{fig:fluxes}~(a) and Fig.~\ref{fig:Tex}.
They well reproduce the $T_\mathrm{ex}$ behavior predicted by the $\textit{ab initio}$ calculation.

\subsubsection{Temperature boundary}

\begin{figure}
    \centering
    \includegraphics[width=7.5cm]{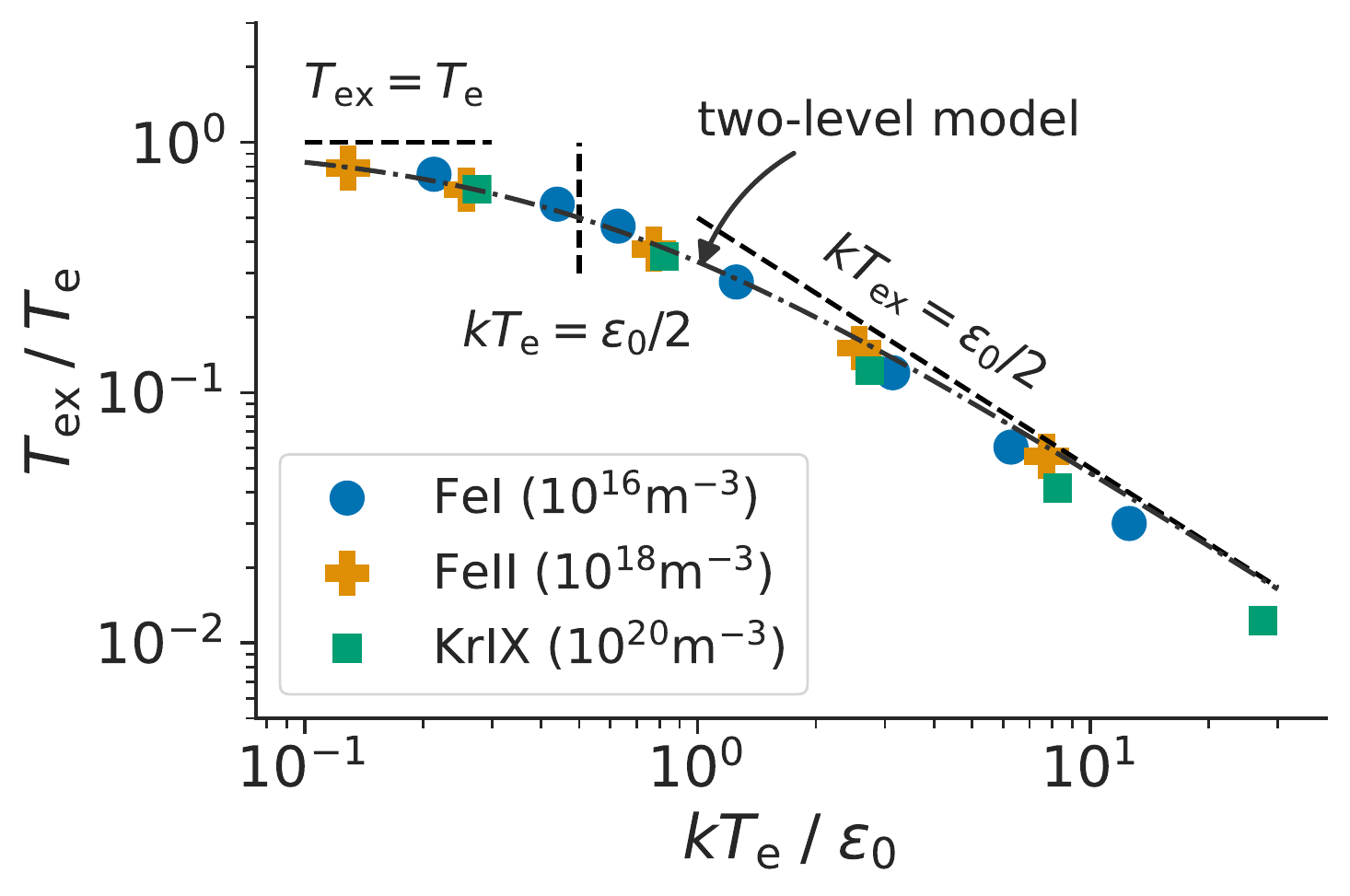}
    \caption{
        $T_\mathrm{e}$-dependence of $T_\mathrm{ex}$ in low-density plasmas, which are simulated using FAC.
        In the high-temperature region, $k T_\mathrm{ex} \approx \epsilon_0 / 2$ is satisfied, and $T_\mathrm{ex} \approx T_\mathrm{e}$ is a good estimate in the low-$T_\mathrm{e}$ region, for all three different many-electron ions.
        The chain curve indicates the limiting behavior derived from two-level model \eref{eq:Teff_2lev_limitting} with $\Delta E = \frac{3}{2}\epsilon_0$..
    }
    \label{fig:temperature_boundary}
\end{figure}

\begin{figure}
    \centering
    \includegraphics[width=7.5cm]{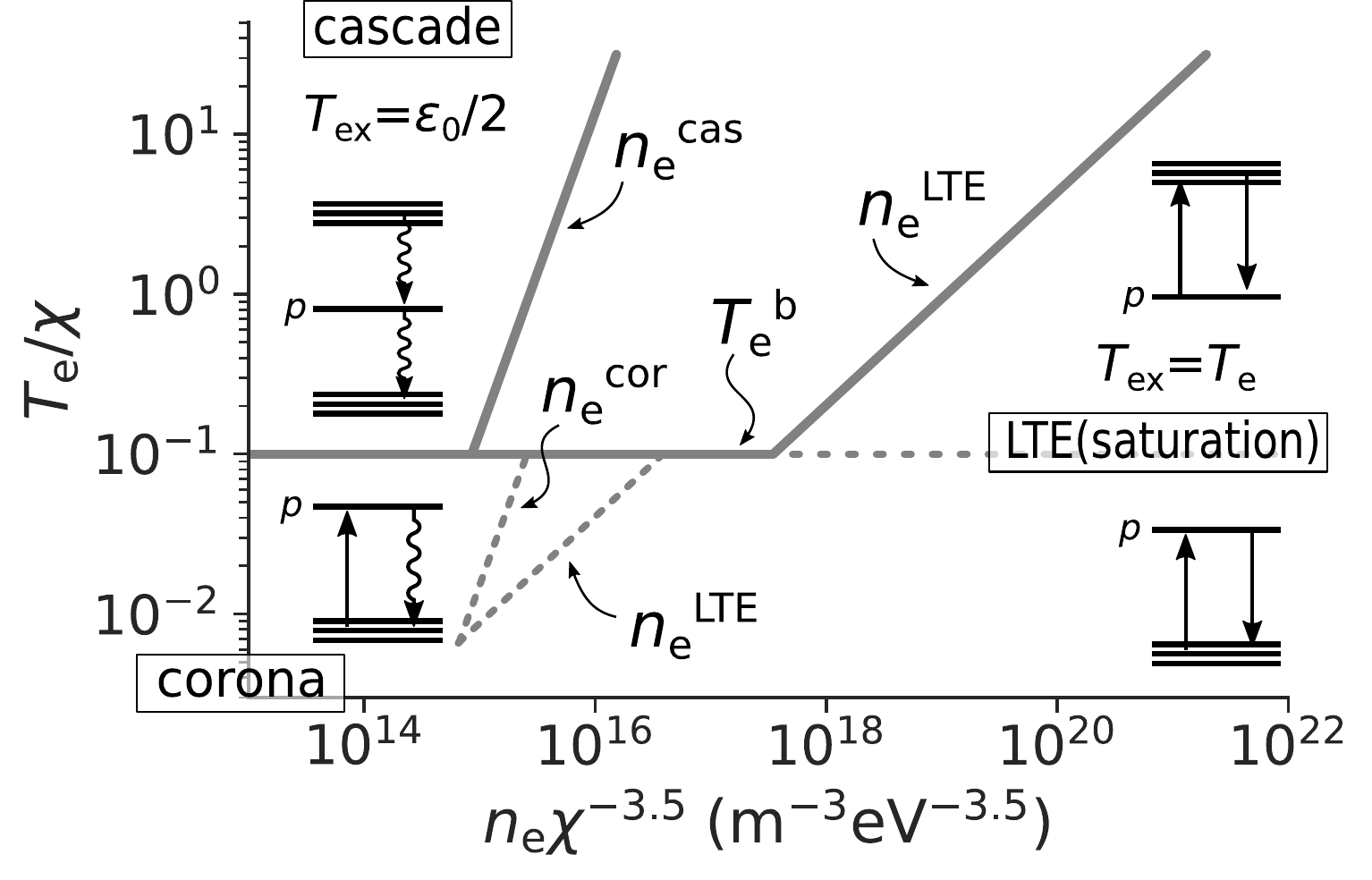}
    \caption{
        Phase diagram for the population kinetics of many-electron atoms.
        Four typical regions, corona, cascade, and LTE phases are depicted.
        Here, we assume $\epsilon_0 \approx 0.2 \chi$ and $\sigma \approx 0.1 \chi$.
    }
    \label{fig:temperature_diagram}
\end{figure}

In Fig.~\ref{fig:temperature_boundary}, we show $T_\mathrm{ex} / T_\mathrm{e}$ for the low-density region computed from the FAC results, as a function of $k T_\mathrm{e} / \epsilon_0$.
The $n_\mathrm{e}$ values and atomic elements are shown in the figure.
All results are on the same curve.
In the figure, we also show the lines $T_\mathrm{ex} = T_\mathrm{e}$ and $k T_\mathrm{ex} = \epsilon_0 / 2$.
It can be seen that these are good estimates of $T_\mathrm{ex}$ in the low- and high-$T_\mathrm{e}$ regions, respectively.
Therefore, the temperature boundary is,
\begin{equation}
    \label{eq:temperature_boundary}
    T_\mathrm{e}^\mathrm{b} = \frac{\epsilon_0}{2k}.
\end{equation}
A diagram illustrating these phases and their boundaries is shown in Fig.~\ref{fig:temperature_diagram}.

\subsection{Low Excited States\label{subsec:low_excited_state}}

\begin{figure*}
    \centering
    \includegraphics[width=10cm]{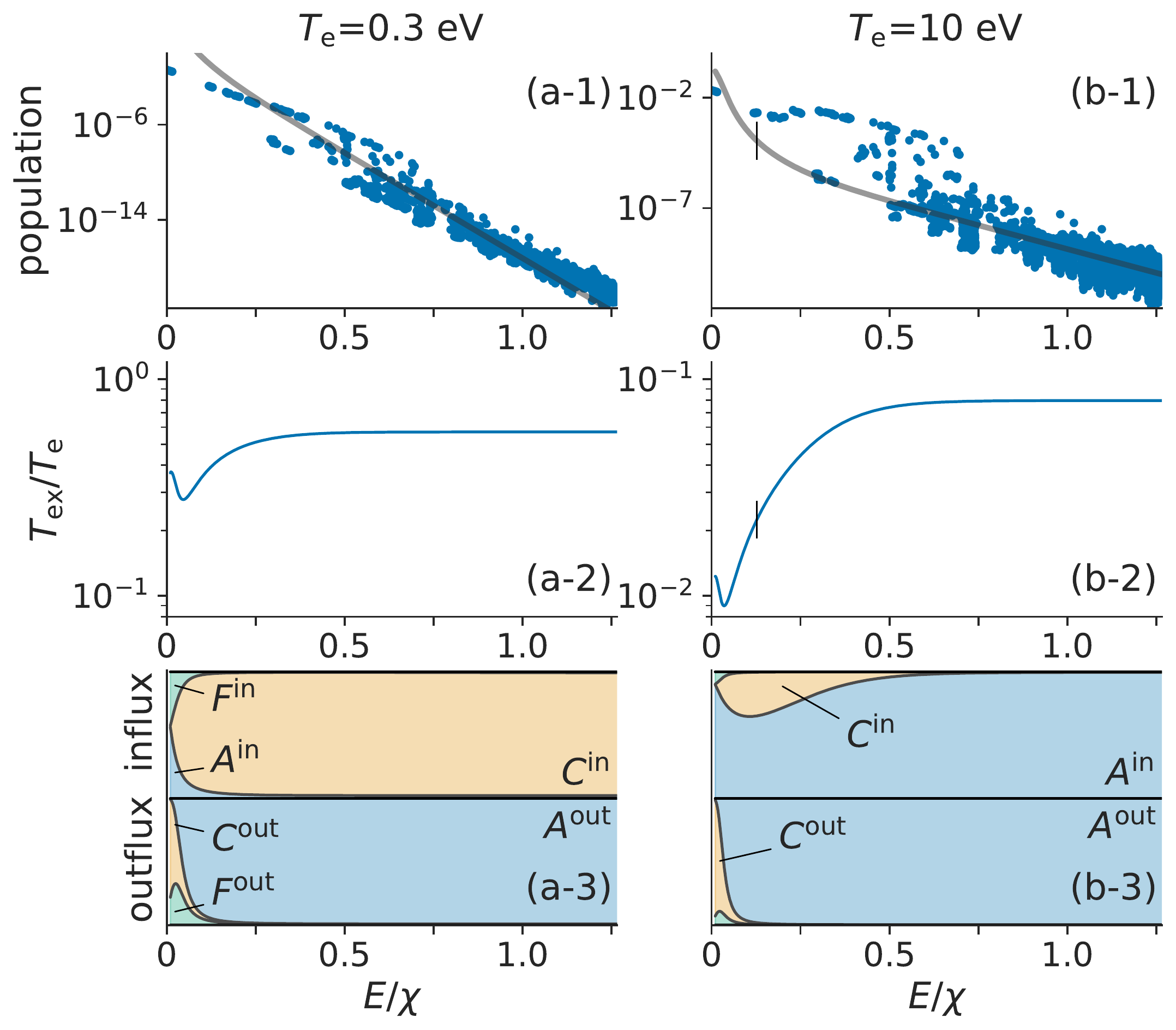}
    \caption{
        Excited energy dependence of the population kinetics for Fe\textsc{I} with $n_\mathrm{e} = 10^{16} \mathrm{\;(m^{-3})}$ and (a) $T_\mathrm{e} = 0.3$ eV and (b) $T_\mathrm{e} = 10$ eV.
        (a-1) and (b-1) show the population distributions computed using the FAC (markers) and our continuous model.
        (a-2) and (b-2) show the local excitation temperature computed from the derivative of the model result.
        (a-3) and (b-3) show the flux compositions as functions of the excited temperature $E$.
        In low-temperature plasmas, the constant $T_\mathrm{ex}$ is reasonable, whereas in high-temperature plasmas, $T_\mathrm{ex}$ significantly drops in low excited states because of the contribution of electron-impact excitation.
    }
    \label{fig:energy_dependence}
\end{figure*}

In the discussion above, we have focused on highly excited states.
In this section, we discuss the energy dependence of the population kinetics.
In Fig.~\ref{fig:energy_dependence}~(a-1) and (b-1), we show the population distribution of Fe\textsc{I} computed using the FAC (markers) and our continuous model (lines) with $T_\mathrm{e}$ = 0.3 eV and 10 eV, respectively, and $n_\mathrm{e} = 10^{16} \mathrm{\;m^{-3}}$.
Both results of FAC and the CCRM deviate from a Boltzmann-like distribution \eref{eq:asymptotic} in the low-energy region ($E < 5$ eV for the FAC and $E < 2.5$ eV for CCRM).
This deviation is clearer in the high-temperature case.

The observed inconsistency between FAC and our model is also more prominent in the low-energy region.
This indicates that in this energy range the assumptions that we use to derive our model, i.e., sufficiently large level density and wavefunction mixing, are not valid.
For example, as there are only even-parity levels in Fe\textsc{I} at $E < 2$ eV, other even levels at higher excited energy cannot decay to these levels by electric-dipole transitions. 
This situation is far different from what we consider in Section \ref{sec:model}.
Although the applicability is limited, for completeness, here we discuss the population kinetics in low-energy levels based on our continuous CR model.

In Fig.~\ref{fig:energy_dependence}~(a-2) and (b-2), we show the energy dependence of $kT_\mathrm{ex}(E) = -n(E) \big/ \frac{\mathrm{d}n(E)}{\mathrm{d}E}$.
In low-temperature plasmas, $T_\mathrm{ex}$ is almost constant for all excited energies, whereas the drop in $T_\mathrm{ex}$ in the low-energy region is significant in high-temperature plasmas.
The flux decomposition is shown in Fig.~\ref{fig:energy_dependence}~(a-3) and (b-3) as a function of the excited energy.

In high-temperature plasmas, the dominant influx and outflux are both spontaneous decay except for $E < 0.5$ eV.
However, if $E \lesssim \delta$, the approximation of \eref{eq:infinte_approximation} is not valid, and thus \eref{eq:Aout_highE} either.
If $E \approx 0$, \eref{eq:Aout} can be approximated as follows, from the Taylor expansion of $e^{-E / \mu}$ up to the fourth order:
\begin{equation}
    \mathcal{A}^\mathrm{out} \approx \gamma \rho_0 S_0 e^{E / \tau} E^4 / 4.
\end{equation}
From $\mathcal{A}^\mathrm{out} = \mathcal{A}^\mathrm{in}$, the excitation temperature may be approximated by
\begin{equation}
    kT_\mathrm{ex} \approx \frac{1}{\frac{24^{1/4}}{E} - \frac{1}{\delta}},
\end{equation}
where $T_\mathrm{ex}$ has $E$ dependence.

As in highly excited states, $k T_\mathrm{ex} \approx \epsilon_0 / 2$, the boundary energy may be defined at $\frac{1}{\frac{24^{1/4}}{E} - \frac{1}{\delta}} = \epsilon_0 / 2$.
This gives the boundary energy,
\begin{equation}
    E^\mathrm{b} = 24^{1/4} \mu
\end{equation}
This boundary energy is shown in Fig.~\ref{fig:energy_dependence}~(b-1) and (b-2) by the vertical bars.

In low-temperature plasmas, the violation of the infinite-range-integration approximation \eref{eq:infinte_approximation} becomes significant also for the influx, i.e., electron-impact excitation.
The decrease in the influx due to the boundary effect compensates for the decrease in the outflux, and therefore, the change in $T_\mathrm{ex}$ is smaller than that in high-temperature plasmas.

\subsection{Relation to the two-level model}
The effective temperature, in particular that of less chaotic systems, has been estimated based on the two-level model, where the population balance of only two levels are considered \cite{Hansen2006}.
In this subsection, we compare our CCRM with the two-level model.

We consider a two-level system consisting of one of excited levels in an many-electron atom (the upper state) and the ground state (the lower state).
Let $n(\Delta E)$ and $n_0$ be the population in the upper and lower states in a two-level system, respectively, while $\Delta E$ be the energy interval between them.
Under our approximations of the rates (Eqs. (\ref{eq:radiative_rate}) and (\ref{eq:excitation_rate})), $n(\Delta E)$ can be written as follows,
\begin{equation}
    n(\Delta E) = n_0 e^{-\frac{\Delta E}{kT_\mathrm{e}}} 
    \left(1 + \xi\Delta E^3 \right)^{-1},
\end{equation}
with
\begin{equation}
    \xi = \frac{\gamma \sqrt{kT_\mathrm{e}}}{\beta n_\mathrm{e}}.
\end{equation}
The effective temperature for the upper state may be represented by,
\begin{align}
    \label{eq:Teff_2lev}
    kT_\mathrm{ex}(\Delta E) 
    & = -\frac{n(\Delta E)}{\frac{\mathrm{d}n(\Delta E)}{\mathrm{d}\Delta E}}\\
    & = 
    kT_\mathrm{e} \left(
        1 + \frac{3 \xi \Delta E^2}{1 + \xi \Delta E^3}
        kT_\mathrm{e}
    \right)^{-1}.
\end{align}
With this model, the limiting behavior of the effective temperature can be written as
\begin{align}
    \label{eq:Teff_2lev_limitting}
    \frac{T_\mathrm{ex}}{T_\mathrm{e}} \rightarrow
    \begin{cases}
        1 & (T_\mathrm{e} \rightarrow 0) \\
        \frac{\Delta E}{3 kT_\mathrm{e}}  & (T_\mathrm{e} \rightarrow \infty) \\
    \end{cases}
\end{align}

For many-electron atoms, it is not obvious whether this model is applicable or what value of $\Delta E$ is appropriate.
However, from the comparison between \eref{eq:Tex_highT_lowN} and \eref{eq:Teff_2lev_limitting}, we find that
\begin{equation}
    \Delta E = \frac{3}{2} \epsilon_0
\end{equation}
gives a consistent limiting behavior.
The $n_\mathrm{e}$ dependence of $T_\mathrm{ex}$, which is computed by the two-level model is shown by chain curves in Fig. \ref{fig:Tex}.
This is similar to our CCRM as well as the simulation results by FAC.

In the limit of $n_\mathrm{e} \rightarrow 0$, \eref{eq:Teff_2lev} gives
\begin{align}
    \frac{T_\mathrm{ex}}{T_\mathrm{e}} \rightarrow
    \left(
        1 + \frac{2 kT_\mathrm{e}}{\epsilon_0}
    \right)^{-1}.
\end{align}
This dependence is shown in Fig. \ref{fig:temperature_boundary} by a chain curve. 
This agrees the first-principle calculation very well.

The similarity between the two-level system and our CCRM is surprising as the two-level-system assumes very different physics from the population kinetics in many-electron atoms.
For example, in low density and high temperature plasmas, the dominant populating flux is the radiative cascade from the upper levels and the dominant depopulating flux is to the lower levels.
This behavior cannot be accounted by two-level systems, as obviously more than two levels are involved.
However, very similar behavior of $T_\mathrm{ex}$ are obtained from the two-level system with only one tuning parameter, $\Delta E$.
This suggests a common underlying mathematical structure that determines the $T_\mathrm{ex}$-behavior as well as $\Delta E = \frac{3}{2}\epsilon_0$, which may be investigated in future studies.

\subsection{Limitations of the CCRM}

In our CCRM, we neglect some important processes, such as electron-impact ionization, auto-ionization, radiative recombination and dielectronic recombination.
In optically thick plasmas, the radiative excitation, radiative recombination and stimulated decay are also important.

From the comparison with the reference FAC simulation, which includes the population outflux by electron-impact ionization and auto-ionization processes, it is found that their effect is negligible within the accuracy of our discussion.
This may be understood from the difference in the level density between charge states.
As can be seen from \eref{eq:epsilon_scaling} the level density is smaller in the next charge state.
There are much less auto-ionization paths (and electron-impact ionization paths) than those of radiative decay (and electron-impact excitation, respectively).
Therefore, even if the rates of auto-ionization and electron-impact ionization are larger, their total contribution to the population distribution can be smaller.

On the other hand, the radiative recombination and dielectronic recombination are not included either in the FAC or our CCRM.
In order to consider their effect, the ion abundance in the next charge state is necessary.
Statistical treatment of the ionization equilibrium is the scopes of the future study.

With the external radiation field, radiative processes, such as radiative excitation, ionization and stimulated radiative decay become important.
If the radiation field can be approximated by a gray-body, these processes can be included in \eref{eq:crm} in a straightforward manner as their rates are simply proportional to the line strength.
However, if the radiation field has a different spectral dependence, it becomes more difficult to generalize.
The study for such a case is also left for the future study.

\begin{figure*}
    \centering
    \includegraphics[width=13.5cm]{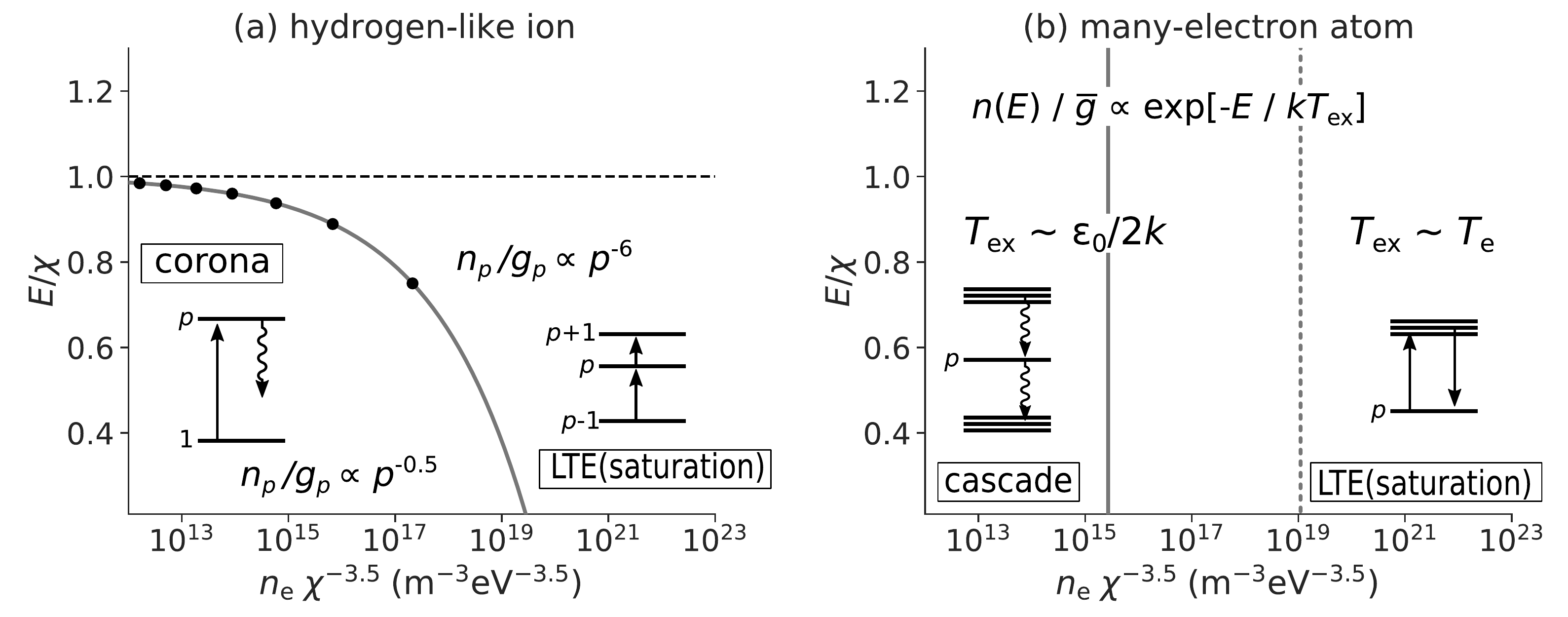}
    \caption{
        Diagrams illustrating the population kinetics of (a) H-like ions studied by Fujimoto~\cite{Fujimoto} (see also Appendix \ref{sec:hydrogen}) and (b) a many-electron atom, which is investigated in this work, in ionizing plasmas.
        The electron temperature is set to $T_\mathrm{e} = \chi / k$, where $\chi$ is the first ionization energy.
        The lower density side in (a) is called the corona phase, where the dominant population process of an excited state is  electron-impact excitation from the ground state, and the depopulation process is spontaneous decay.
        The higher density region is called the LTE phase (or saturation phase), where the dominant population and depopulation processes are electron-impact excitation from the next lower level and to the next higher level, respectively. The population distribution in this region becomes $n_p/g_p\propto p^{-6}$ (Eq.~\ref{eq:minus_sixth}) where $p$ indicates the principal quantum number of H-like ions. The density boundary \eref{eq:boundary_hydrogen} is shown by the gray line.
        On the higher density side in (b), the population distribution is close to the Boltzmann distribution.
        The dominant population process is de-excitation from higher levels, and the dominant depopulation process is the excitation to higher levels.
        The lower density side in (b) is called the cascade phase, where the dominant population process is radiative decay from higher levels, and the dominant depopulation process is radiative decay to lower levels.
        Even in the lower density region, the population distribution of many-electron atoms is similar to the Boltzmann distribution, but with the excitation temperature $T_\mathrm{ex} \approx \epsilon_0 / 2k$, which is independent of $T_\mathrm{e}$ and $n_\mathrm{e}$.
        The density boundary (\eref{eq:ne_cascade}) is shown as a solid vertical line.
        The dashed vertical line indicates another density boundary based on the excitation temperature, where $T_\mathrm{ex} = 0.9 T_\mathrm{e}$ (\eref{eq:ne_sat_highT}) is satisfied.
    }
    \label{fig:diagram}
\end{figure*}

\section{Summary}
In this work, we studied the population kinetics of many-electron atoms in plasmas.
From the statistical theory of the many-electron-atom structure, we constructed a continuous CR model that has only two atom-specific energy scales as parameters, $\epsilon_0$ and $\sigma$.
From this model, the population distribution in highly excited states was found to be Boltzmann-like, but with excitation temperature sometimes smaller than the electron temperature.
We also clarified that there are different phases depending on values of $n_\mathrm{e}$ and $T_\mathrm{e}$ and derived analytical representations of the boundaries.

Some of our findings can be directly used for plasma diagnostics.
For example, the Boltzmann method has been frequently used to estimate $T_\mathrm{e}$, based on the slope of the population distribution and the assumption of the saturation phase, and therefore, the applicability of this method has been limited only to high-density plasmas.
However, as can be seen in Fig.~\ref{fig:Tex}, if $T_\mathrm{ex} < \epsilon_0 / 2k$, then $T_\mathrm{ex} \approx T_\mathrm{e}$ can be inferred.
This clearly shows much wider applicability of the Boltzmann method for low-temperature plasmas.
This property also enables us to use a new temperature diagnostics using line intensity statistics, which has been proposed in Ref.~\cite{Fujii2020}.
By contrast, if $T_\mathrm{ex} \gtrsim \epsilon_0 / 2k$, then the inference of $T_\mathrm{e}$ may be difficult without knowing $n_\mathrm{e}$.

For highly charged ions in low-density and high-temperature plasmas, such as heavy ions in tokamak core plasmas or in electron-beam ion traps, the population is mostly concentrated in the low excited states, to which our model is not applicable.
However, our finding \eref{eq:Tex_highT_lowN} may still be useful to estimate the cascade contribution from very highly excited states, which is difficult to consider from first-principles, since there are infinite number of possible levels.

In this work, we only compared our model with another simulation model, FAC.
Comparison with experimental observation is desirable; however, because of the difficulty in the level identification and accurate computation of the transition rates for highly excited states, it is not available at the current stage.
We leave it for future studies.

In principle, our CCRM could be further developed to include additional atomic structure data, such as more sophisticated line-strength functions, based on individual orbitals within the statistical theory of many-body quantum chaos~\cite{Flambaum1998}. While this would not add significant computational overhead, the simplicity of our current formulation, \eref{eq:strength-function}, allows for analytical exploration of the effective excitation temperature through phase space.

\begin{acknowledgments}
This work was partly supported by JSPS KAKENHI Grant Number 19K14680, the grant of Joint Research by the National Institutes of Natural Sciences (NINS program No, 01111905), and partly by the Max-Planck Society for the Advancement of Science. JCB is supported by the Alexander von Humboldt Foundation. We thank Jos\'e Crespo Lop\'ez-Urrutia for useful discussions.
\end{acknowledgments}

\appendix

\section{Population Kinetics of H-like Ions in Plasmas\label{sec:hydrogen}}

The study of CR models was started from the simplest atoms, H- and He-like ions \cite{griem_1997,Fujimoto,Sawada1993,Goto2003,Fujimoto1979}. It was expanded to analyze more complex ions and molecules as detailed calculation of the atomic and molecular structure became available.

For H and other simple ions such as He or alkali metals, the population kinetics has been understood systematically by Griem and Fujimoto et al. \cite{Fujimoto, Griem}.
As shown in Fig.~\ref{fig:diagram}~(a), they clarified the existence of typical two phases for ionizing plasmas.
The analytical representation of the boundary has been also studied.
Here, we briefly summarize their works.

The line strength of an H-like atom is known to have the following asymptotic dependence \cite{Menzel1935,Bethe1957},
\begin{equation}
    \label{eq:line_strength_hydrogen}
    S_{pq} = \frac{2^6}{\sqrt{3}\pi} p^{-3} q^{-3} 
    (p^{-2} - q^{-2})^{-4} z^{-2} e^2 a_0^2 g_\mathrm{bb}
\end{equation} 
where $z$ is the nuclear charge of the H-like ion and $g_{bb}$ is the bound--bound gaunt factor, which is close to 1.
Although in \eref{eq:crm_each_term}, $p$ and $q$ should represent any states, and they are not necessarily quantum numbers, only in this subsection, we assume $p$ and $q$ to correspond to the principal quantum numbers for H-like ions.

In the low-density limit of ionizing plasmas, the population is more concentrated on the ground state, because of the strong population flow to lower levels by spontaneous transitions.
Therefore, collisional excitation from the ground state is the dominant population process for excited states.
In this case, $\mathcal{C}^\mathrm{in}$ and $\mathcal{A}^\mathrm{out}$ can be written as follows:
\begin{align}
    \notag
    \mathcal{C}^\mathrm{in} 
    & \approx C_{p \gets 1} n_\mathrm{e} n_1 \\
    \notag
    & \approx \frac{1}{g_1}\frac{\beta n_\mathrm{e}}{\sqrt{kT_\mathrm{e}}} \exp\left[-\frac{\omega_{p1}}{kT_\mathrm{e}} \right]S_{1p}\\
    \label{eq:Cin_hydrogen}
    & \approx \frac{1}{2}\frac{\beta n_\mathrm{e}}{\sqrt{kT_\mathrm{e}}}
    p^{-3} z^{-2} 
    \exp\left(-\frac{\omega_{p1}}{kT_\mathrm{e}}\right)
    \frac{2^6}{\sqrt{3}\pi} e^2 a_0^2
\end{align}
and
\begin{align}
    \notag
    \mathcal{A}^\mathrm{out} 
    & \approx \sum_{q < p} A_{q \gets p} n_p \\
    \notag
    & \approx \sum_{q < p} \gamma \frac{1}{g_p} \omega_{pq}^3 S_{pq} n_p \\
    \label{eq:Aout_hydrogen}
    & \approx 0.7\cdot \frac{3}{2^5} \gamma E_\mathrm{H}^3 p^{-4.5} z^{-4} \frac{2^6}{\sqrt{3}\pi} e^2 a_0^2 n_p
\end{align}
Here, we substituted \eref{eq:line_strength_hydrogen} into \eref{eq:radiative_rate} and \eref{eq:excitation_rate} and assumed $p \gg 1$ and $\log p \approx 0.7 p^{0.5}$\cite{Fujimoto}.
From the flux balance $\mathcal{C}^\mathrm{in} \approx \mathcal{A}^\mathrm{out}$, the population at state $p$ can be written as follows:
\begin{align}
    \label{eq:minus_half}
    n_p / g_p \propto p^{-0.5}
\end{align}

In high-density plasmas, the dominant population process becomes electron-impact excitation from the next lower level, and the dominant depopulation process is electron-impact excitation to the next higher level,
\begin{align}
    C_{p+1 \gets p} n_\mathrm{e} n_p 
    \approx C_{p \gets p-1} n_\mathrm{e} n_{p-1} 
    \approx \mathrm{(const.)}.
\end{align}
This leads $ n_p \propto (C_{p+1 \gets p})^{-1}$, and thus \cite{Fujimoto}
\begin{align}
    \label{eq:minus_sixth}
    n_p / g_p \propto p^{-6}.
\end{align}
In this phase, the population distribution does not depend on $n_\mathrm{e}$.
We call this the saturation phase.

The transition boundary between the corona and saturation phases may be defined when the dominant outflux changes from spontaneous transition to electron-impact excitation,
\begin{align}
    \sum_{q<p} A_{q\gets p} \approx C_{p + 1 \gets p} n_\mathrm{e},
\end{align}
which leads to \cite{Fujimoto1973,Fujimoto,Griem} \eref{eq:boundary_hydrogen}.
A schematic diagram of the population kinetics is shown in Fig.~\ref{fig:diagram} (a).

\section{FAC Computation\label{sec:fac_config}}
In this section, we show the detailed setup for the atomic structure calculation using the FAC.

The FAC utilizes the configuration interaction method, where a many-body wavefunction is approximated by a linear combination of single-body wavefunctions.
In principle, more basis wavefunctions give better accuracy.
Table \ref{sup:tb:configuration} shows the configurations used to simulate the atomic structure of Fe\textsc{I}, Fe\textsc{II}, and Kr\textsc{IX}.
The total numbers of the basis functions for Fe\textsc{I}, Fe\textsc{II}, and Kr\textsc{IX} are 5427, 6997, and 3489, respectively.

\begin{table}[h!]
    \caption{List of configurations to simulate the atomic structure using the FAC.
    The notation of $n$*$m$ indicates the use of all possible orbital combinations of $m$ electrons in the shell with the principal quantum number $n$.}
    \label{sup:tb:configuration}
    \centering
    \begin{tabular}{c | c | c} 
    \hline
    Kr\textsc{IX} & Fe\textsc{I} & Fe\textsc{II}\\
    \hline
    $3d^{10}$ & $3d^8$ & $3d^7$\\
    $3d^8$ 4*2 & $3d^6 4s^2$ & $3d^5 4s^2$\\
               & $3d^6 4s 4p$ & $3d^5 4s 4p$ \\
               & $3d^6 4s 4d$ & $3d^5 4s 4d$ \\
               & $3d^6 4s 4f$ & $3d^5 4s 4f$ \\
    $3d^8 4s$ 5*1 & $3d^6 4s$ 5*1 & $3d^5 4s$ 5*1 \\
    $3d^9 4s$ 4*1 & $3d^7$ 4*1 & $3d^6$ 4*1 \\
               & $3d^7$ 5*1 & $3d^6$ 5*1 \\
    \hline
    \end{tabular}
\end{table}

The configuration interaction method should converge to the true value if we include infinite number of basis wavefunctions with any arbitrary central potential.
However, in reality, we may need to tune the potential.
We tune them, so that some of the computed low-lying levels and the ionization potential match to those data compiled in NIST ASD.

\bibliography{refs} 


\end{document}